\documentclass[letterpaper,twocolumn,10pt]{article}
\usepackage{usenix,epsfig,endnotes}

\usepackage{graphicx}
\usepackage{cite}
\usepackage{url}
\usepackage{color}
\usepackage[dvipsnames]{xcolor}
\usepackage{alltt}
\usepackage{hyperref} %change this later
\hypersetup{
	citebordercolor={0 1 0},
	citecolor=blue,
	colorlinks=false,
	filebordercolor={0 .5 .5},
	filecolor=green,
	linkbordercolor={1 0 0},
	linkcolor=blue,
	menubordercolor={1 0 0},
	pageanchor=true,
	pagebackref=true,
	pdfborder={0 0 1},
	pdfpagelabels=true,
	pdftex,
	plainpages=false,
	urlbordercolor={0 1 1},
	urlcolor=blue,
}
\usepackage{syntax}
\usepackage{newfloat}
\usepackage{multicol}

\usepackage[labelfont=bf, skip=0pt]{caption}
\newcommand{\Comment}[1]{}

\newcommand{\SmallSpace}{\vspace*{-1.5ex}}

\newcommand{\pgao}{\textcolor{red}}

\newcommand{\myparatight}[1]{\smallskip\noindent{\bf {#1}:}}

\DeclareFloatingEnvironment[
fileext   = logr,% don't use log here!
listname  = {List of Grammars},
name      = Grammar,
placement = !htbp
]{BNF}

\newcommand{\distance}{5pt}
\usepackage[ruled,vlined, noend]{algorithm2e}
\usepackage{mathtools}
\usepackage{url}
\usepackage{multirow}
\usepackage{mdwlist}
\usepackage{xspace}
\usepackage{amsmath}
\usepackage{amssymb}
\usepackage{misc}
\usepackage{subcaption}
\usepackage{listings}
\usepackage{enumerate}

\usepackage{enumitem}
\usepackage{adjustbox}
\usepackage{mdframed}

\definecolor{mygreen}{rgb}{0,0.6,0}
\definecolor{mygray}{rgb}{0.5,0.5,0.5}
\newcommand{\incode}[1]{\lstinline{#1}}

\newcommand{\eat}[1]{}

\newcommand{\dsl}{\textsc{Aiql}\xspace}
\newcommand{\dslff}{\textsc{Aiql}\_FF\xspace}

\newcommand{\code}[1]{\emph{#1}}

\newif \ifcomments
\commentstrue

\ifcomments
    \newcommand{\kjee}[1]{{-\textcolor{red}{#1}-}}
\else
    \newcommand{\kjee}[1]{}
\fi

\newcommand{\eg}{e.g., }
\newcommand{\ie}{i.e., }

\SetKwInput{KwInput}{Input}
\SetKwInput{KwOutput}{Output}
\SetKwProg{Fn}{Function}{}{end}

\SetCommentSty{mycommfont}

\begin{document}
\date{}

%make title bold and 14 pt font (Latex default is non-bold, 16 pt)
\title{\Large \bf \dsl: Enabling Efficient Attack Investigation \\ from System Monitoring Data}

\author{
{\rm Peng Gao$^1$}
\and
{\rm Xusheng Xiao$^2$}
 \and
 {\rm Zhichun Li$^3$}
 \and
 {\rm Kangkook Jee$^3$}
 \and
 {\rm Fengyuan Xu$^4$}
 \and
 {\rm Sanjeev R. Kulkarni$^1$}
 \and
 {\rm Prateek Mittal$^1$}
 \and
{\normalsize $^1$Princeton University\; $^2$Case Western Reserve University\; $^3$NEC Laboratories America, Inc.}
\and
{\normalsize $^4$National Key Lab for Novel Software Technology, Nanjing University}
 \and
{\small $^1$\{pgao,kulkarni,pmittal\}@princeton.edu\; $^2$xusheng.xiao@case.edu\; $^3$\{zhichun,kjee\}@nec-labs.com\; $^4$fengyuan.xu@nju.edu.cn} 
}

\maketitle

% Use the following at camera-ready time to suppress page numbers.
% Comment it out when you first submit the paper for review.
\thispagestyle{empty}
\pagestyle{empty}

%%%%%%%%%%%%%%%%%%%%%%%%%%%%%%%%%%%%%%%%%%%%%%

\subsection*{Abstract}
The need for countering Advanced Persistent Threat (APT) attacks has led to the solutions that ubiquitously monitor system activities in each host,
and perform timely attack investigation over the monitoring data for analyzing attack provenance.
However, existing query systems based on relational databases and graph databases lack language constructs to express key properties of major attack behaviors,
and often execute queries inefficiently since their semantics-agnostic design cannot exploit the properties of system monitoring data to speed up query execution.

To address this problem, we propose a novel query system built on top of existing monitoring tools
and databases, which is designed with novel types of optimizations to support timely attack investigation.
Our system provides (1) domain-specific \emph{data model and storage} for scaling the storage,
(2) a domain-specific query language, \emph{Attack Investigation Query Language (\dsl)} that integrates critical primitives for attack investigation, and
(3) an optimized query engine based on the characteristics of the data and the semantics of the queries to efficiently schedule the query execution.
We deployed our system in NEC Labs America comprising 150 hosts and evaluated it using 857 GB of real system monitoring data (containing 2.5 billion events).
%To evaluate how our system supports timely attack investigation,
%we measure both the efficiency in query execution time and the conciseness in query specification.
Our evaluations on a real-world APT attack and a broad set of attack behaviors	
show that our system surpasses existing systems in both efficiency (124x over PostgreSQL, 157x over Neo4j, and 16x over Greenplum)
and conciseness (SQL, Neo4j Cypher, and Splunk SPL contain at least 2.4x more constraints than \dsl).

%%%%%%%%%%%%%%%%%%%%%%%%%%%%%%%%%%%%%%%%%%%%%%
\section{Introduction}

Advanced Persistent Threat (APT) attacks are sophisticated (involving many individual attack steps across many hosts and exploiting various
vulnerabilities) and stealthy (each individual step is not suspicious enough), plaguing many well-protected businesses~\cite{tc,ebay, opm, homedepot, target, equifax}.
%A recent TARGET data breach~\cite{target} caused a net earning loss of 1.2 billion dollars.
A recent massive Equifax data breach~\cite{equifax} has exposed the sensitive personal information of 143 million US customers. 
%Unlike conventional attacks, these advanced attacks are sophisticated (involving many individual attack steps across many hosts and exploiting various software vulnerabilities) and stealthy (each individual step is not suspicious enough)~\cite{tc}.
%Thus, enterprises have a strong need for solutions to ``connect the suspicious dots'' across multiple activities.
%This requires (1) large-scale collection and storage of ``the suspicious dots'' (attack provenance)
%and (2) attack investigation that ``connects the dots'' for identifying risky system behaviors.
%
In order for enterprises to counter advanced attacks,
%it is crucial to understand the activities of hosts at a fine-grained level. 
recent approaches based on \emph{ubiquitous system monitoring} have emerged as an important solution for monitoring system activities and performing attack investigation~\cite{backtracking,backtracking2,taser,taserdb, intrusionrecovery,mpi,Ma:2015:ALC:2818000.2818039,lee2013high}. 
System monitoring observes \emph{system calls} at the kernel level to collect system-level events about system activities.
%providing a comprehensive way to capture system behaviors across all hosts in the enterprises.
Collection of system monitoring data enables security analysts to investigate these attacks by 
\emph{querying risky system behaviors} over the historical data~\cite{securitylandslide}.
%from historical data~\cite{securitylandslide} 
%by \emph{querying risky system behaviors} over the system monitoring data.

Although attack investigation is performed after the attacks compromise enterprises' security, it is a considerably time-sensitive task due to two major reasons. 
First, advanced attacks include a sequence of steps and are performed in multiple stages. 
%A detected suspicious behavior may not be the very end of an attack sequence and the attacks could cause more damage if not successfully contained.  
A timely attack investigation can help understand all attack behaviors and prevent the further damage of the attacks.
Second, understanding the attack sequence is crucial to correctly patch the systems.
A timely attack investigation can pinpoint the vulnerable components of the systems and protect the enterprises from future attacks of the same types.

%In the life cycle of attack investigation, 
%security analysts play an important role from intrusion identification to compromise clean up.

\myparatight{Challenges} However, there are two major challenges for building a query system to support security analysts in efficient and timely attack investigation.
%these attack investigations.
\begin{figure*}[!ht]
	%\vspace{-0.1cm}
	\centering
	\includegraphics[width=0.95\textwidth]{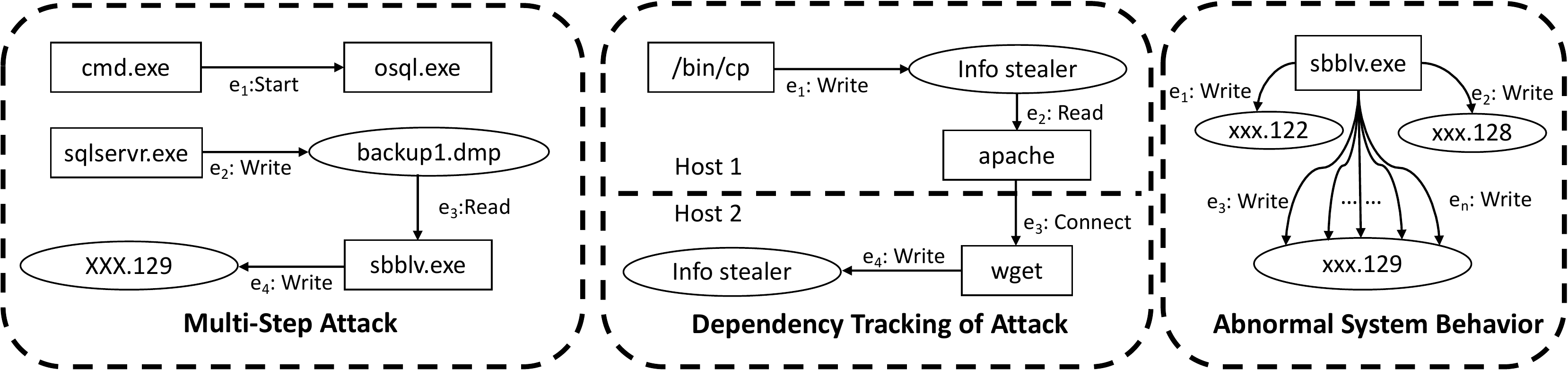}
	\vspace{1ex}
	\caption{Major types of attack behaviors (events $e_1,\ldots,e_n$ are shown in ascending temporal order)}
	\label{fig:moti}
%\vspace*{-1ex}
\end{figure*}

\emph{Attack Behavior Specification}: The system needs to provide a query language with specialized constructs for expressing various types of attack behaviors using system monitoring data:
(1) \textbf{Multi-Step Attacks}: risky behaviors in advanced attacks typically involve activities that are related to each other based on either specific attributes (\eg the same process reads a sensitive file and accesses the network) or temporal relationships (\eg file read happens before network access), which requires language constructs to easily specify \emph{relationships among activities}.
In Fig.~\ref{fig:moti}, the attacker runs \incode{osql.exe} to cause the database \incode{sqlservr.exe} to dump its data into a file \incode{backup1.dmp}. Later (i.e., \incode{e3} happens after \incode{e2}; temporal relationship), a malicious script \incode{sbblv.exe} reads from the dump \incode{backup1.dmp} (i.e., the same dump file in \incode{e2} and \incode{e3}; attribute relationship) and sends the data back to the attacker. 
(2) \textbf{Dependency Tracking of Attacks}: dependency analysis is often applied to track causality of data for discovering the ``attack entry'' (\ie provenance)~\cite{backtracking,backtracking2,backtrackingfile},
which requires language constructs to \emph{chain constraints among activities}.
In Fig.~\ref{fig:moti}, a malicious script \incode{info_strealer} in Host 1 infects Host 2 via \emph{network communications between \incode{apache} and  \incode{wget}}.
(3) \textbf{Abnormal System Behaviors}: frequency-based behavioral models are often required to express abnormal system behaviors, such as network access spikes~\cite{netspike, adware}.
Investigating such spikes requires the system to support \emph{sliding windows and statistical aggregation} of system activities, and compare the aggregate results with either \emph{fixed thresholds (in absolute sense)} or the \emph{historical results (in relative sense)}.
In Fig.~\ref{fig:moti}, a malicious script \incode{sbblv.exe} sends a \emph{large} amount of data to a particular destination \incode{XXX.129}.\footnote{While existing complex event processing systems~\cite{flink,esper,siddhi} support similar features,
they operate over stream rather than historical data stored in databases.}

\emph{Big-Data Security Analysis}: System monitoring produces a huge amount of daily logs~\cite{loggc,reduction} ($\sim$ 50 GB per day for 100 hosts),
and the investigation of these attacks typically requires enterprises to keep at least a $0.5\sim 1$
%half year to one 
year worth of data~\cite{trustwave}.
Such \emph{a big amount of security data} poses challenges for the system to meet the requirements of \emph{timely} attack investigation.

\emph{Limitations of Existing Systems}:
Unfortunately, existing query systems do not address both of these inherent challenges in attack investigation.
First, existing query languages in relational databases based on SQL and SPARQL~\cite{postgresql,sql,sparql} lack constructs for easily chaining constraints among relations.
Graph databases such as Neo4j~\cite{neo4j} and NoSQL tools such as MongoDB~\cite{chodorow2013mongodb}, Splunk~\cite{splunk}, and ElasticSearch~\cite{elasticsearch} are ineffective in expressing event relationships where two events have no common entities (\eg $e_1$ and $e_2$ in Fig.~\ref{fig:moti}).
More importantly, none of these languages provide language constructs to express behavioral models with historical results.
%Additionally, for the behaviors that can be expressed by these languages, the resulting queries usually have many constraints, making them laborious and error-prone to compose and debug.
%For example, a SQL query for the risky behavior that consists of three events and three relationships requires about 30 constraints (described in \red{Query}~\ref{inv:a1:ext}).
%For example, a SQL query (Query~\ref{c4-8:sql} in for expressing an APT attack step consisting of 7 events and 6 relationships (behavior described in \dsl Query~\ref{c4-8:stail}) requires 77 constraints and 2792 characters , and a Neo4j Cypher query (Query~\ref{c4-8:cypher}) requires 63 constraints and 2570 characters.
Second, system monitoring data is generated with a timestamp on a specific host in the enterprise, exhibiting strong \emph{spatial and temporal properties}. 
However, none of these systems provide optimizations that exploit the domain specific characteristics of the data,
missing opportunities to optimize the system for supporting timely attack investigation and often causing queries to run for hours
% rather than seconds 
(e.g., performance evaluation results in Sec.~\ref{case:eval-results}).

\myparatight{Contributions}
We design and build a novel system for efficient attack investigation from system monitoring data.
%We build our system on top of existing relational databases such as PostgreSQL~\cite{postgresql}, 
%and can be further scaled up using parallel processing databases such as Greenplum~\cite{greenplum},
%which enable our system to leverage the services provided by these mature infrastructures, such as data management, recovery, and security.
We build our system ($\sim$ 50,000 lines of Java code) on top of existing 
system-level monitoring tools (\ie auditd~\cite{auditd} and ETW~\cite{etw}) for data collection and relational databases (\ie PostgreSQL~\cite{postgresql} and Greenplum~\cite{greenplum}) for data storage and query. 
This enables our system to leverage the services provided by these mature infrastructures, such as data management, indexing mechanisms, recovery, and security.
In particular, our system is designed with three novel types of optimizations.
First, our system provides a domain-specific query language, \emph{Attack Investigation Query Language (\dsl)},
which is optimized to express the three aforementioned types of attack behaviors.
Second, our system provides domain-specific \emph{data model and storage} for scaling the storage.
Third, our system optimizes the query engine based on the characteristics of the monitoring data and the semantics of the queries to efficiently schedule the query execution.
To the best of our knowledge, we are \emph{the first to accelerate attack investigation via optimizing storage and query of system monitoring data}.

\begin{lstlisting}[captionpos=b, caption={\dsl Query for CVE-2010-2075~\cite{cve1}}, label={inv:a1:ext}]
agentid = 1 // host id; spatial constraints
(at "01/01/2017") // temporal constraints
proc p1 start proc p2["%telnet%"] as evt1
proc p3 start ip ipp[dstport = 4444] as evt2
proc p4["%apache%"] read file f1["/var/www%"] as evt3
with p2 = p3,  // attribute relationship
evt1 before evt2, evt3 after evt2 // temporal relationships
return p1, p2, p4, f1
\end{lstlisting}
%\vspace*{-1ex}

\begin{figure*}[!ht]
	%\vspace{-0.1cm}
	\centering
	\includegraphics[width=0.95\textwidth]{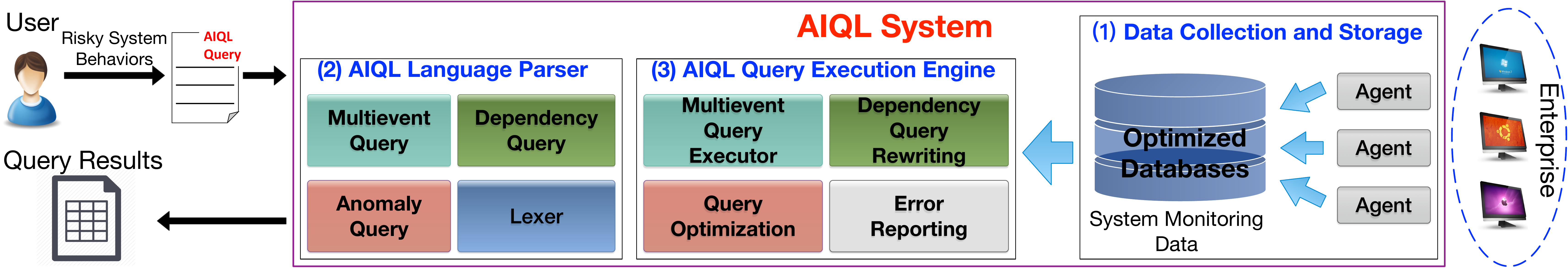}
	\caption{The \dsl system architecture}
	\label{fig:architecture}
%	\vspace*{-1ex}
\end{figure*}

\emph{Domain-Specific Query Language} (Sec.~\ref{sec:design}):
Our \dsl language is designed for specifying the attack behaviors shown in Fig.~\ref{fig:moti} 
(\ie Query~\ref{case:stail:c5:comp} in Sec.~\ref{case:investigation}, Query~\ref{inv:d2} in Sec.~\ref{subsec:path}, and Query~\ref{case:stail:anomaly} in Sec.~\ref{case:investigation}, respectively).
Specifically, \dsl provides language constructs to specify relationships among system activities (Sec.~\ref{subsec:multievent}),
chain constraints among activities (Sec.~\ref{subsec:path}), and compute aggregate results in sliding time windows (Sec.~\ref{subsec:anomaly}).
\dsl adopts the \emph{\{subject-operation-object\}} syntax to represent system behavior patterns as events (\eg \incode{proc p1 write file f1}) and supports \emph{attribute relationships} and \emph{temporal relationships} of multiple events, as well as syntax shortcuts based on context-aware inference (Sec.~\ref{subsec:multievent}). 
As shown in Query~\ref{inv:a1:ext}, \dsl can relate multiple system activities using spatial/temporal constraints and attribute/temporal relationships.
%The explicit constructs for spatial/temporal constraints and relationships allow the query engine to explicitly optimize the query search over spatial and temporal dimensions.
%In contrast to previous languages such as SQL~\cite{sql}, SPARQL~\cite{sparql}, Cypher~\cite{cypher}, and Splunk~\cite{splunk}, \dsl's novelty lies in the integration of critical primitives for attack investigation, 
%including explicit spatial and temporal constraints, relationship specifications, constraint chaining among system activities, 
%and access to aggregate and historical results in sliding time windows.

\emph{Data Model and Storage} (Sec.~\ref{subsec:datastorage}):
Our system models system monitoring data as a sequence of events,
where each event describes how a process interacts with a system resource, such as writing to a file. 
More importantly, our system clearly identifies the spatial and temporal properties of the events,
and leverages these properties to partition the data storage in both \emph{spatial and temporal dimensions}.
%Distinguishing system entities enables better data deduplications in storing system monitoring data while preserves the accuracy of attack investigation.
%which enables domain-specific optimizations such as automatic space-time partitioning for splitting the data into chunks. 
%Data deduplication specific to system entities in system monitoring data reduces the storage pressures in storing the already seen system entities.
%Automatic space-time partitioning exploits the temporal and spatial properties of system monitoring data to distribute the data storage:
%(1) \textbf{spatial partition}: it stores system monitoring data in partitions, where the data of a specific host is stored in a specific partition, 
%allowing faster accesses for queries that search for the data of specific hosts;
%(2) \textbf{temporal partition}: it further partitions the data along the time dimension, 
%allowing faster accesses for queries that search for the data in a specific period.
Such partitioning presents opportunities for parallel processing of query execution
% processing when querying over the data 
(Sec.~\ref{subsec:optimization}).

\emph{Query Scheduling} (Sec.~\ref{sec:engine}):
Our system identifies both \emph{spatial} and \emph{temporal} constraints in \dsl queries, and optimizes the query execution in two aspects:
(1) for \dsl queries that involve multiple event patterns,
% for searching related activities, 
our system prioritizes the search of event patterns with high pruning power, 
maximizing the reduction of irrelevant events as early as possible;
(2) our system breaks down the query into independent sub-queries along temporal and spatial dimensions and executes them in parallel.

%%%%%%%%%%%%%%%%%%
\myparatight{Evaluation}
%We built the \dsl system ($\sim$ 50,000 lines of Java code) upon existing system-level monitoring tools (i.e., auditd~\cite{auditd} and ETW~\cite{etw}) and deployed it in an enterprise system comprising around 100 hosts.
We deployed the \dsl system in NEC Labs America comprising 150 hosts.
We performed a broad set of attack behaviors in the deployed environment, and evaluated the query performance and %query 
conciseness of \dsl against existing systems using 857 GB of real system monitoring data (16 days; 2.5 billion events):
(1) our end-to-end efficiency evaluations on an APT attack case study (27 queries) show that \dsl surpasses both PostgreSQL (124x)  and Neo4j (157x);
(2) our performance evaluations show that the query scheduling employed by \dsl is efficient in both single-node databases (40x over PostgreSQL scheduling) and parallel 
%processing 
databases (16x over Greenplum scheduling);
(3) our conciseness evaluations on four major types of attack behaviors (19 queries) show that SQL, Neo4j Cypher, and Splunk SPL contain at least 2.4x more constraints, 3.1x more words, and 4.7x more characters than \dsl.
%In addition to the conciseness and efficiency, our evaluations also demonstrate \dsl's expressiveness in a broad set of attack behaviors.
All 
%the evaluation 
queries and a demo video are available on our \emph{project website~\cite{aiql}}.

\section{System Overview and Threat Model}
\label{sec:overview}
% Goal: query subgraphs
% Data model: RDBMS, temporal events, three types, relationships
%We build a ubiquitous enterprise-wide audit system to collect and store the system
%monitoring data, and implement a query engine to efficiently search the system monitoring data.
Fig.~\ref{fig:architecture} shows the \dsl system architecture:
%, which consists of three components.
%data collection, TBQL language parser, and TBQL query execution engine.
%\myparatight{Data Collection and Storage}
(1) we deploy monitoring agents across servers, desktops and laptops in the enterprise
to monitor system activities by collecting information about system calls from kernels.
The collected system monitoring data
%from each host 
is then sent to the central server and stored in our optimized 
data storage (Sec.~\ref{sec:model-storage});
%databases, which is optimized using domain specifics, for efficient data storage and retrieval.
%\myparatight{TBQL Language Parser}
(2) the language parser, implemented using ANTLR 4~\cite{antlr}, 
%performs syntactic and semantic analysis of 
analyzes input queries and generates query contexts. A query context is an object abstraction of the input query that contains all the required information for the query execution. 
Multievent syntax, dependency syntax, and anomaly syntax are supported (Sec.~\ref{sec:design});
(3) the query execution engine executes the generated query contexts to search for the desired attack behaviors. 
Based on the data storage and the query semantics, domain-specific optimizations, such as relationship-based scheduling and temporal $\&$ spatial parallelization, are adopted to speedup the query execution (Sec.~\ref{sec:engine}).
%for query scheduling are adopted.

\eat{
The query engine executes TBQL queries to find desired behaviors.
The multievent query executor is the core module to interpret the generated parse tree and search the events accordingly.
The dependency query rewriting module % is a language extension module, which
compiles the original dependency query into an equivalent query represented using the core grammar.
The query optimization module leverages domain-specific characteristics of the system monitoring data to optimize the query execution.
The error reporting module monitors the query execution, and reports errors in different execution stages.
}

\myparatight{Threat Model}
Our thread model follows the threat model of previous work~\cite{backtracking,backtracking2,loggc, lee2013high,trustkernel}. 
%Particularly, we define the trusted computing base (TCB) for causality analysis to be the kernel mechanisms, 
%the backend database that stores and manages system monitoring logs. 
We assume that kernel is trusted, and the system monitoring data collected from kernel 
%space 
is not tampered with~\cite{auditd,etw}.
Any kernel-level attack that deliberately compromises security auditing systems is beyond the scope of this work.
%We do consider that insiders or external attackers have full knowledge of ``normal'' activities, so that they can intentionally launch attacks with seemingly normal operations and may poison our reference database using a burst of repeated malicious activities.

%%%%%%%%%%%%%%%%%%%%%%%%%%%%%%%%%%%%%%%%%%%%%%
\section{Data Model and Storage}
\label{sec:model-storage}

\subsection{Data Model and Collection}
\label{subsec:datamodel}
%\myparatight{Data Model}
System monitoring data records the interactions among system resources as system events~\cite{backtracking}. 
Each of the recorded event occurs on a particular host at a particular time, thus exhibiting strong spatial and temporal properties.
%Each event can naturally be described as a system entity (subject) does some operation on another system entity (object).
Existing works have indicated that on most modern operating systems (Windows, Linux and OS X), system resources (system entities) in most cases are files, processes, and network connections~\cite{backtracking,backtracking2,taser,wormlog}. 
Thus, in our data model, we consider system \emph{entities} as \emph{files}, \emph{processes}, and \emph{network connections}. 
We define a system event as the interaction among two system entities represented using the triple \emph{$\langle$subject, operation, object$\rangle$}, which consists of the initiator of the interaction, the type of the interaction, and the target of the interaction. 
Subjects are processes originating from software applications such as Firefox, and objects can be files, processes and network connections.
We categorize system events into three types according to their object entities, namely \emph{file events}, \emph{process events}, and \emph{network events}.

Both entities and events have critical security-related attributes (Tables~\ref{tab:entity-attributes} and~\ref{tab:event-attributes}).
The attributes of entities include the properties to support various security analyses (\eg file name, process name, and IP addresses), 
and the unique identifiers to distinguish entities (\eg file data ID and process ID).
%For a file entity, we use volume id and data id (including inode id and file igeneration number\footnote{Could be obtained via \incode{ioctl} in Linux.}) as its unique identifier. 
%For a process entity, we use process id, process starting time, and an ordinal number\footnote{Processes that start at the same time are distinguished using this number.} as its unique identifier. 
%For a network connection entity, as processes usually communicate with some servers using different network connections but with the same IPs and ports, treating these connections differently greatly increases the amount of data we trace and such granularity is not required in most of the cases. Thus, we use 5-tuple (\emph{$\langle$srcip, srcport, dstip, dstport, protocol$\rangle$)} as a network connection's unique identifier. 
%Failing to distinguish different entities causes problems in relating events to entities, especially for tracking dependencies of events, which makes the system ineffective in querying precise information.
The attributes of events include event origins (\ie agent ID and start time/end time),
operations (\eg file read/write),
and other security-related properties (\eg failure code).
Agent ID refers to the unique ID of the host where the entity/event is observed. 

\begin{table}[t]
	\centering
	\caption{Representative attributes of system entities}\label{tab:entity-attributes}
	\begin{adjustbox}{width=0.4\textwidth}
		\begin{tabular}{|l|l|}
			\hline
			\textbf{Entity}		&\textbf{Attributes}\\\hline
			File				&Name, Owner/Group, VolID, DataID, etc.\\\hline
			Process			&PID, Name, User, Cmd, Binary Signature, etc.\\\hline
			Network Connection	& IP, Port, Protocol \\\hline
		\end{tabular}
	\end{adjustbox}
	%	\vspace*{-2ex}
	
	%	\vspace*{-2ex}
\end{table}

\myparatight{Data Collection}
We implement data collection agents for Windows and Linux
based on ETW event tracing~\cite{etw} and the Linux Audit Framework~\cite{auditd}.
Tables~\ref{tab:entity-attributes} and~\ref{tab:event-attributes} show representative attributes of our collected data. %(agent ID refers to the unique ID of the host where the entity/event is collected from).

\subsection{Data Storage}
\label{subsec:datastorage}
After the modeling, we store the data in relational databases powered by PostgreSQL~\cite{postgresql}.
Relational databases come with mature indexing mechanisms and are scalable to massive data.
However, even with indexes for speeding up queries, relational databases still face challenges in handling high ingest rates of massive 
system monitoring data.
We next describe how we address these challenges to optimize the database storage.

\myparatight{Time and Space Partitioning}
System monitoring data exhibits strong \emph{temporal and spatial properties}:
the data collected from different agents is independent from each other,
and the timestamps of the collected data increase monotonically.
Queries of the data are often specified with a specific time range or a host, or across many hosts within some time interval.
Therefore, when storing the data, we partition the data in both the time and the space dimensions:
separating groups of agents into table partitions and generating one database per day for the data collected on that day. 
%To query an event, users need to join three tables: subject table, object table, and event table.
We build various types of indexes on the attributes that will be queried frequently,
such as executable name of process, name of file, source/destination IP of network connection.

\myparatight{Hypertable}
For large organizations with hundreds or thousands of machines, we scale the data storage using MPP (massively parallel processing) databases Greenplum~\cite{greenplum}.
These databases intelligently distribute the storage and search of events and entities based on the spatial and temporal properties of our data model.

% unique identifiers into several database instances,

%which can then be deployed in multiple servers.
%Queries in MPP databases leverage the distributed structure and parallelism to speed up searches.

%%%%
\myparatight{Time Synchronization}
We correct potential time drifting of events on agents by applying synchronization protocols like Network Time Protocol (NTP)~\cite{ntp} at the client side, and checking with the clock at the server side. %(Appendix~\ref{appendix:time-sync}). 
%Details are in Appendix~\ref{appendix:time-sync}.

\begin{table}[t]
	\centering
	\caption{Representative attributes of system events}\label{tab:event-attributes}
	\begin{adjustbox}{width=0.4\textwidth}
		\begin{tabular}{|l|l|}
			\hline
			Operation		& Read/Write, Execute, Start/End, Rename/Delete\\\hline
			Time/Sequence		& Start Time/End Time, Event Sequence\\\hline
			Misc.		& Subject ID, Object ID, Failure Code\\\hline
		\end{tabular}
	\end{adjustbox}
	%	\vspace*{-2ex}
	
	%	\vspace*{-4ex}
\end{table}

\section{Query Language Design}
\label{sec:design}

\dsl is designed to specify three types of attack behaviors: multi-step attacks, dependency tracking of attacks, and abnormal system behaviors.
In contrast to previous query languages~\cite{sql,sparql,cypher,splunk}
%such as SQL, SPARQL, Cypher, and Splunk 
that focus on the specification of relation joins or graph paths,
%paths in a graph,
\dsl uniquely integrates the critical primitives for attack investigation,
providing explicit constructs for spatial/temporal constraints, relationship specifications, constraint chaining among system events, 
and the access to aggregate and historical results in sliding time windows.
Grammar~\ref{bnf:parser} shows the representative rules of \dsl.

%%%%%%%%%%%%%%%%%%%%%%%
\subsection{Multievent \dsl Query}
\label{subsec:multievent}
For multievent queries, \dsl provides explicit language constructs for system events (in a natural format of \emph{\{subject-operation-object\}}), spatial/temporal constraints, and event relationships.

\myparatight{A Running Example}
%Multievent \dsl query specifies one or more event patterns and their relationships. 
Query~\ref{query:multi-history} specifies an example 
%risky 
system behavior that probes user command history files.
%: \incode{.viminfo} and \incode{.bash_history},
Multiple context-aware syntax shortcuts (illustrated in comments) are used, such as attribute inference and omitting unreferenced entity IDs (details are given later).
%Multiple syntax shortcuts are used, such as \emph{context-aware attribute inferences} (illustrated in comments) and omitting unreferenced entity IDs. 

%We will use this query as a running example to illustrate different features of \dsl.
\eat{
Query~\ref{query:history} 
%shows one risky behavior expressed in \dsl 
specifies one event pattern
that probes user command history files:
\incode{.viminfo} and \incode{.bash_history}.
%\footnote{We will use this query as a running example to illustrate different features of \dsl.}. 

\begin{lstlisting}[captionpos=b, caption={Command history probing}, label={query:history}]
proc p read file f[name = ".viminfo" || name = ".bash_history"] as evt
return p.exe_name
\end{lstlisting}
}
%A multi-event query defines multiple event patterns. 
%Query~\ref{query:multi-history} extends Query~\ref{query:history} by adding the process creation event \incode{p2 start p1}. 

\begin{lstlisting}[captionpos=b, caption={Command history probing}, label={query:multi-history}]
agentid = 1 // unique id of the enterprise host
(at "01/01/2017") // time window
proc p2 start proc p1 as evt1
proc p3 read file[".viminfo" || ".bash_history"] as evt2 // .viminfo -> name=.viminfo; omit file ID
with p1 = p3, evt1 before evt2
return p2, p1 //p2 -> p2.exe_name, p1 -> p1.exe_name
sort by p2, p1
\end{lstlisting}
%\vspace*{-2ex}

%%%%%%%%%%%%%%%
\myparatight{Global Constraints}
The global constraint rule (\emph{$\langle$global_cstr$\rangle$}) specifies the constraints for all event patterns (e.g., \code{agentid} and \code{time window} in Query~\ref{query:multi-history}).

%%%%%%%%%%%%%%%
%\vspace*{-3ex}
\myparatight{Event Pattern}
The event pattern rule (\emph{$\langle$evt_patt$\rangle$}) specifies an event pattern that consists of the subject/object entity (\emph{$\langle$entity$\rangle$}), operation (\emph{$\langle$op_exp$\rangle$}), and optional event ID (\emph{$\langle$evt$\rangle$}). 
The entity rule (\emph{$\langle$entity$\rangle$}) consists of entity type,
% (file, process, network connection), 
optional entity ID, and optional attribute constraints (\emph{$\langle$attr_cstr$\rangle$}). Logical operators (``$\&\&$'' for AND, ``$||$'' for OR, ``!'' for NOT) can be used in \emph{$\langle$op_exp$\rangle$} and \emph{$\langle$attr_cstr$\rangle$} to form complex expressions. The optional time window rule (\emph{$\langle$twind$\rangle$}) further narrows down the search for the event pattern. Common time formats (US formats and ISO 8601) and granularities are supported.
%For example, \incode{proc p (read || write) file f["name=.viminfo" || "name=.bash_history"] as evt (at "01/01/2017")} specifies on day \incode{01/01/2017}, a process entity \incode{p} reads from or writes to a file entity \incode{f}, whose name can be \incode{.viminfo} or \incode{.bash_history}. 

\begin{BNF}[h]
\footnotesize

\begin{mdframed}

\setlength{\grammarparsep}{-2pt} % increase separation between rules
\setlength{\grammarindent}{8em} % increase separation between LHS/RHS 

\begin{grammar}
<aiql> ::= <multievent> | <dependency>

<multievent> ::= (<global_cstr>)* (<m_query>)+

<dependency> ::= (<global_cstr>)* <d_query>

<global_cstr> ::= <cstr> | `(' <twind> `)' | <slide_wind>

<twind> ::= `from' <datetime> `to' <datetime> | ...

<slide_wind> ::= <wind_length> <wind_step>
\end{grammar}

\textbf{Multi-event query:}
\begin{grammar}
%<m_query> ::= (<query_id> `:')? <evt_patt>+ <evt_rel>? <return> <filter>?
<m_query> ::= <evt_patt>+ <evt_rel>? <return> <filter>?

<evt_patt> ::= <entity> <op_exp> <entity> <evt>? (`(' <twind> `)')?

<entity> ::= <entity_type> <e_id>    ? (`[' <attr_cstr>`]')?

<attr_cstr> ::= <cstr> \alt `!'<attr_cstr>
\alt <attr_cstr> (`&&' | `||') <attr_cstr>
\alt `(' <attr_cstr> `)'

<cstr> ::= <attr> <bop> <val>
\alt `!'? <val>
\alt <attr> `not'? `in' `('  <val> (`,' <val>)* `)'
%\alt <attr> `not'? `in' <query_id>

<op_exp> ::= <op> \alt `!'<op_exp>
\alt <op_exp> (`&&' | `||') <op_exp>
\alt `(' <op_exp> `)'

<evt> ::= `as' <evt_id> (`[' <attr_cstr>`]')?

<evt_rel> ::= `with' <rel> (`,' <rel>)*

<rel> ::= <attr_rel> | <temp_rel>

<attr_rel> ::= <e_id>`.'<attr> <bop> <e_id>`.'<attr>
\alt <e_id> <bop> <e_id>

<temp_rel> ::= <evt_id> (`before' | `after' | `within') (`[' <val>`-'<val> <timeunit>`]')? <evt_id>

<return> ::= `return' `count'? `distinct'? <res> (`,' <res>)*
%, count'?
%\alt `list' `('  <res> (`, ' <res>)* `)'

<res> ::= <e_id>(`.'<attr>)?
\alt <agg_func>`(' <res> `)'
\alt `as' <rename_id>

<group_by> ::= `group by' <res> (`,' <res>)* 

<filter> ::= `having' <expr>
\alt `sort by' <attr> (`,' <attr>)* (`asc' | `desc')?
\alt `top' <int>
\end{grammar}

\textbf{Dependency query:}
\begin{grammar}
<d_query> ::= ((`forward' | `backward') `:')? (<entity> <op_edge>)+ <entity> <return> <filter>?

<op_edge> ::= (`->' | `<-') `[' <op_exp> `]'

\end{grammar}

\end{mdframed}

\vspace*{-2ex}
\caption{Representative BNF grammar of \dsl}\label{bnf:parser}
\end{BNF}
\myparatight{Event Attribute and Temporal Relationships}
\label{subsubsec:event-rels}
The event relationship rule (\emph{$\langle$evt_rel$\rangle$}) specifies how multiple event patterns are related.
The attribute relationship rule (\emph{$\langle$attr_rel$\rangle$}) uses attribute values of event patterns to specify their relationships. In Query~\ref{query:multi-history}, \incode{p1=p3} (inferred as \incode{p1.id=p3.id})
%; details in Sec.~\ref{subsubsec:syntactic-sugar}) 
indicates that two event patterns \incode{evt1} and \incode{evt2} are linked by the same entity. 
The temporal relationship rule (\emph{$\langle$temporal_rel$\rangle$}) specifies temporal order (``before'', ``after'', ``within'') of event patterns. For example, \incode{evt1 before[1-2 minutes] evt2} specifies that \incode{evt1} occurred 1 to 2 minutes before \incode{evt2}.

\myparatight{Event Return and Filters}
The event return rule (\emph{$\langle$return$\rangle$}) retrieves the attributes of the matched events. Constructs such as ``count'', ``distinct'', ``top'', ``having'', and ``sort by'' are provided for result manipulation and filtering.
\myparatight{Context-Aware Syntax Shortcuts}
\dsl includes language syntax shortcuts to make queries more concise.
\begin{itemize}[noitemsep, topsep=1pt, partopsep=1pt, listparindent=\parindent, leftmargin=*]
%	\item \emph{Context-aware attribute inferences}: 
	\item \emph{Attribute inference}: 
%	default attribute names will be inferred if not specified:
	(1)  default attribute names will be inferred if users specify only attribute values in an event pattern, or specify only entity IDs in the return clause. We select the most commonly used attributes in security analysis as the default attributes: \incode{name} for files, \incode{exe_name} for processes, and \incode{dst_ip} for networks;
	% connections; 
	(2) \incode{id} will be used as the default attribute if users specify only entity IDs in attribute relationships.
	
	\item \emph{Optional ID}: the ID of entity/event can be omitted if it is not referenced in the event relationship clause or the event return clause.
	
	\item \emph{Entity ID reuse}: reusing entity IDs in multiple event patterns implicitly means that these event patterns share the same entity.
\end{itemize}
For example, in Query~\ref{query:multi-history}, \incode{".viminfo"}, \incode{return p2}, and \incode{p1 = p3} will be inferred as \incode{name = ".viminfo"}, \incode{return p2.exe_name}, and \incode{p1.id = p3.id}, respectively. Query~\ref{query:multi-history} also omits the file ID in \code{evt2} since it is not referenced. We can also replace \code{p3} with \code{p1} in \code{evt2} and omit  \incode{p1 = p3}. 
\subsection{Dependency \dsl Query}
\label{subsec:path}

%To address the challenges specified in Sec.~\ref{subsec:moti:dep}, 
\dsl provides the dependency syntax that chains constraints and specifies temporal relationships among event patterns, facilitating the specification of dependency tracking of attacks.
The syntax specifies a sequence of event patterns in the form of a path, where nodes in the path represent system entities and edges represent operations.
The \incode{forward} and \incode{backward} keywords can be used to specify the temporal order of the events on the path: \incode{forward} (\incode{backward}) means the events found by the leftmost event pattern occurred earliest (latest).  
%where each node in the path represents a subject/object entity, and these entities are connected via operation edges specified by the \emph{$\langle$op_edge$\rangle$} rule (e.g., \incode{->[write]}, where the arrow (\incode{->}) points from the subject to the object).

\begin{lstlisting}[captionpos=b, caption={Forward tracking for malware ramification}, label={inv:d2}]
(at "01/01/2017")
forward: proc p1["%/bin/cp%", agentid = 2] ->[write] file f1["/var/www/%info_stealer%"] 
<-[read] proc p2["%apache%"] 
->[connect] proc p3[agentid=3] // tracking across host
->[write] file f2["%info_stealer%"] 
return f1, p1, p2, p3, f2 
\end{lstlisting}
%\vspace*{-2ex}

Query~\ref{inv:d2} shows a forward dependency query in \dsl that investigates the ramification of malware (\incode{info_stealer}), which originates from host $h_a$ (\incode{agentid = 2}) and affects host $h_b$ (\incode{agentid = 3}) through an Apache web server. 
Lines 2-3 specify that \incode{p1} writes to \incode{f1}, and then \incode{f1} is read by \incode{p2}.
Such syntax eliminates the need to repetitively specify the shared entity (i.e., \incode{f1}) in each event pattern.
An example result may show that \incode{p3} is the \incode{wget} process that downloads the malicious script from host $h_b$.
The operation \incode{->[connect]} at Line 4 indicates the search will track dependencies of events across hosts.

%%%%%%%%%%%%%%%%%%%%%%%
\subsection{Anomaly \dsl Query}
\label{subsec:anomaly}
\eat{Security analysts have their domain knowledge about the organizations, and have a strong need to perform statistical aggregation of matched patterns and compare the current aggregation results with either pre-defined threshold values or previous aggregation results. For example, malware outbreaks and hacking attempts can cause spikes in network traffic, and hence there is a need to manipulate the queried results to identify such spikes~\cite{netspike}.%https://www.paessler.com/press/pressreleases/top_5_causes_of_sudden_spikes_in_traffic
}

%To address the challenges specified in Sec.~\ref{subsec:moti:suspicious},
\dsl provides the constructs of \emph{sliding time window} with common aggregation functions (e.g., \incode{count}, \incode{avg}, \incode{sum}) to facilitate the specification of frequency-based system behavioral models.
%the computation of behavior frequencies and comparison to certain thresholds.
Besides, \dsl provides the construct of \emph{history states}, allowing queries to compare frequencies using historical information.
\eat{
\begin{lstlisting}[captionpos=b, caption={Simple moving average for network frequency}, label={query:sma3}]
(at "01/01/2017")
window = 1 min
step = 10 sec
proc p read ip ipp
return p, count(distinct ipp) as freq
group by p
having freq > 2 * (freq + freq[-1] + freq[-2]) / 3
\end{lstlisting}
\vspace*{-1ex}
}

\begin{lstlisting}[captionpos=b, caption={Simple moving average for network frequency}, label={query:sma3}]
(at "01/01/2017")
window = 1 min
step = 10 sec
proc p read ip ipp
return p, count(distinct ipp) as freq
group by p
having freq > 2 * (freq + freq[1] + freq[2]) / 3
\end{lstlisting}
%\vspace*{-1ex}

Query~\ref{query:sma3} shows an anomaly query that specifies a 1-minute sliding time window and computes the moving average~\cite{hamilton1994time} 
%SMA3
to detect network spikes (Line 7).
%Query~\ref{query:sma3} specifies a 1-minute sliding window with sliding step being 10 seconds, and computes the count of distinct network accesses within each sliding window. The aggregation results are then filtered by comparing the current value with twice of its three-period simple moving average (SMA)~\cite{hamilton1994time}, which is computed by averaging the previous aggregation results. Only processes that have a network access spike will be returned.
%To facilitate such computations, 
\dsl supports the common types of moving averages through built-in functions (SMA, CMA, WMA, EWMA~\cite{hamilton1994time}). 
For example, the computation of EWMA for network frequency with normalized deviation can be expressed as: \incode{(freq - EWMA(freq, 0.9)) / EWMA(freq, 0.9) > 0.2}.

%// having amount > 2 * SMA(amount, 3) // Simple moving average
%// having amount > 2 * WMA(amount, 3) // Weighted moving average
%// having amount > 2 * EWMA(amount, 0.9) // Exponential weighted moving average
%// having (amount - EWMA(amount, 0.9)) / EWMA(amount, 0.9) > 0.2 // EWMA with normalized deviation

%%%%%%%%%%%%%%%%%%%%%%%%%%%%%%%%%%%%%%%%%%%%%%
\section{Query Execution Engine}
\label{sec:engine}
%\dslquery execution engine performs syntactic and semantic analysis of \dslqueries,

The \dsl query execution engine executes the query context generated by the parser
and optimizes the query execution by leveraging domain-specific properties of system monitoring data.
Optimizing a query with many constraints is a difficult task due to the complexities of joins and constraints~\cite{sql-tuning}.
\dsl addresses these challenges by providing explicit language constructs for spatial/temporal constraints and temporal relationships,
so that the query engine can directly optimize the query execution by:
(1) using event patterns as a basic unit for generating data queries and leveraging attribute/temporal relationships to optimize the search strategy;
(2) leveraging the spatial and temporal properties of system monitoring data to partition the data and executing the search in parallel based on the spatial/temporal constraints. 

%%%%%%%%%%%%%%%%%%%
\subsection{Query Execution Pipeline}
\label{subsec:pipeline}

Fig.~\ref{fig:multievent} shows the execution pipeline of a multievent query.
%, which is used by the complete subquery execution module in Figure~\ref{fig:subquery}. 
Based on the query semantics, for every event pattern, the engine synthesizes a \emph{SQL data query}, which searches the optimized relational databases (Sec.~\ref{subsec:datastorage}) for the matched events.
The data query scheduler prioritizes the execution of data queries to optimize execution performance (Sec.~\ref{subsec:optimization}). 
Execution results of each data query are further processed by the executor
% (implemented using Java) 
to perform joins and filtering to obtain the desired results.
Note that by weaving all these join and filtering constraints together, the engine could generate a large SQL with many constraints mixed together.
Such strategy suffers from indeterministic optimizations due to the large number of constraints and often causes the %query 
execution to last for minutes or even hours (Sec~\ref{case:eval-results}).
For an input dependency query, the engine compiles it to an equivalent multievent query for execution. 
For an anomaly query, the engine maintains the aggregate results as historical states and performs the filtering based on the historical states.
%The execution of a dependency query is done by compiling it to an equivalent multievent query for execution.

%%%%%%%%%%%%%%%%%%%
\eat{
\myparatight{Query Rewriting}
\label{subsec:exec-path}
Query rewriting translates queries expressed using a language extension to an equivalent multievent query. 
In this work, we provide the syntax of dependency query as the language extension to facilitate dependency tracking (details of translation from a dependency query to a multievent query are in Appendix~\ref{appendix:translation}).
}

%\vspace*{-4ex}

%%%%%%%%%%%%%%%%%%%
\subsection{Data Query Scheduler}
\label{subsec:optimization}
\begin{figure}[t]
	%\vspace{-0.1cm}
	\centering
	\includegraphics[width=0.45\textwidth]{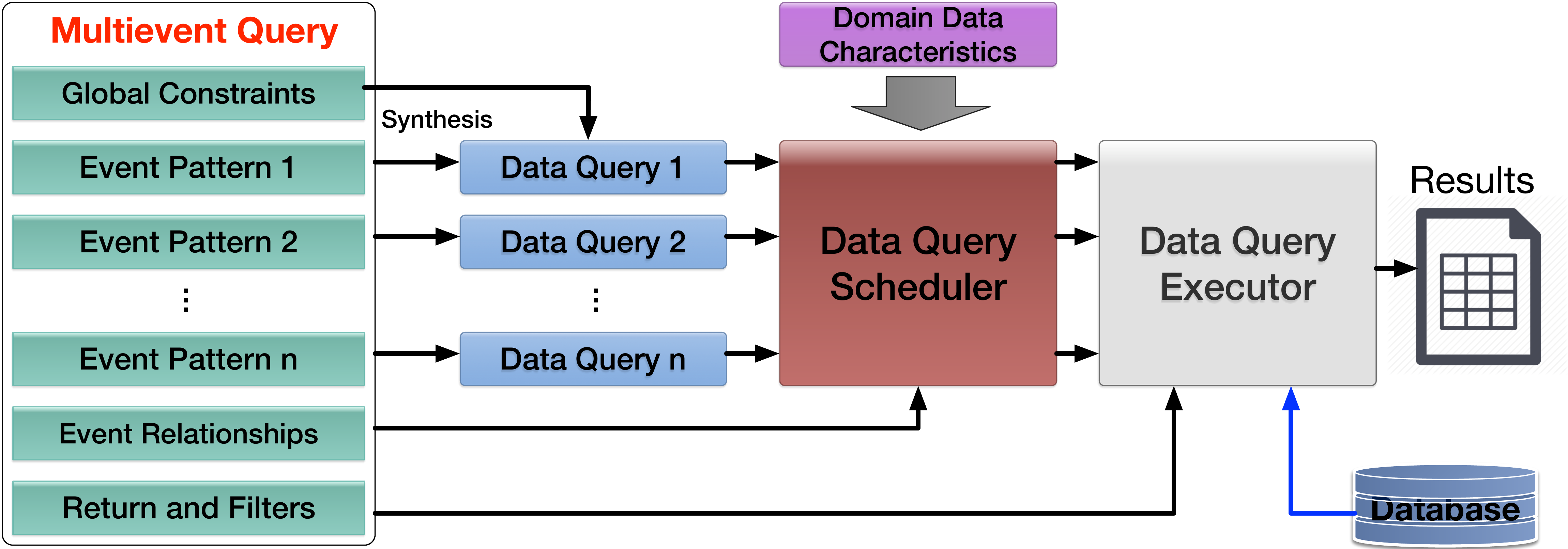}
	\caption{Execution of a multievent \dsl query}
	\label{fig:multievent}
	%\vspace*{-4ex}
\end{figure}

The data query scheduler in Fig.~\ref{fig:multievent} schedules the execution of data queries. 
A straightforward scheduling strategy (\emph{fetch-and-filter}) is to: 
(1) execute data queries separately and store the results of each query in memory;
(2) leverage event relationships to filter out results that do not satisfy the constraints. 
However, this strategy incurs non-trivial computation costs and memory space if some data queries return a large number of results.

\myparatight{Relationship-Based Scheduling}
To optimize the execution scheduling of data queries,
we leverage two insights based on event relationships: (1) event patterns have different levels of pruning power, and the query engine can prioritize event patterns with more pruning power to narrow the search;
% of tuples;
%Instead of data queries first and enumerating all possible tuples and event relationships to do filtering, we can adopt a ``constructive'' approach, by feeding in event relationships with larger pruning power and executing involved, unexecuted data queries on the fly. We maintain sets of tuples of event IDs which satisfy all previously analyzed event relationships, and update or merge tuple sets after analyzing the current event relationship. 
(2) if two event patterns are associated with an event relationship, the query engine can execute the data query for the pattern that has more constraints first (likely having more pruning power), and use the execution results to constrain the execution of the other data query. 

%For example, given the event relationship \code{evt1.dst_id = evt2.src_id}, and $evt1$ is associated with more attributes (hence more pruning power) than $evt2$, we can execute $evt1$ first, and use the return \emph{dst_id} to constrain the search of $evt2$.

Algorithm~\ref{alg:schedule} gives the \emph{relationship-based} scheduling: 
\begin{enumerate}[label={\arabic*.}, noitemsep, topsep=1pt, partopsep=1pt, listparindent=\parindent, leftmargin=*]
	\item A pruning score is computed for every event pattern based on the number of constraints specified.

	\item Event relationships are sorted based on the relationship type (process events and network events are sorted in front of file events) and the sum of the involved event patterns' pruning scores.
	% Relationships with higher sum of scores are sorted in the front.
	%, which indicate the estimated pruning power.

	\item The main loop processes event relationships returned from the sorted list, executes data queries, and generates result tuples. 
	The engine executes the data query whose associated event pattern has a higher pruning score first, and leverages existing results to narrow the search scope.
%	The execution 
%	%of data queries 
%	leverages event patterns with higher pruning scores and stored results to narrow the search scope. 
	To facilitate tuple management, we maintain a map $M$ that stores the mapping from the event pattern ID to the set of event ID tuples that its execution results belong to. 
	As the loop continues, new tuple sets are created and put into $M$, and old tuple sets are updated, filtered, or merged.

	\item After analyzing all event relationships, if there remain unexecuted data queries, these queries are executed and the corresponding results are put into $M$.

	\item The last step is to merge tuple sets in $M$, so that 
	all event patterns are mapped to the same tuple set that satisfy all constraints.
\end{enumerate}
%\vspace*{-2ex}
%%%%%%%%%%%

\begin{algorithm}[!h]
\footnotesize

%\DontPrintSemicolon
\SetAlgoLined
\KwInput{$n$ data queries: $Q = \{q_i \mid i \leq n, i \in \mathbb{N}^+\}$\newline 
$n$ event patterns: $E=\{e_i \mid i \leq n, i \in \mathbb{N}^+\}$\newline
$m$ event relationships: $R = \{rel(e_i, e_j)\}$
}
\KwOutput{Event ID tuples that satisfy all constraints
%: $S_1 \times S_2 \times \cdots S_n \mid _{R}$
%Set of event ID tuples: $S_1 \times S_2 \times \cdots S_n \mid _{R}$. $S_i$ is the set of event IDs after executing  $q_i$, and $\mid _R$ means that every tuple satisfies all event relationships in $R$.
}

%\KwData{data queries , event relationships $\{rel_i\}$}
%\KwResult{set of segments: $segs$}

1. $\forall e_i\in E, score(e_i) \xleftarrow{compute} e_i$\;
%The computation is based on the number of $evt_i$'s attribute constraints.\;

2. $R_{sorted} \xleftarrow{sort} R$\;
%The sorting is based on relationship type and scores of related event patterns. Relationships with more pruning power will be sorted in the front.\;

3. Initialize empty set $Exec$, empty map $M$\;
\For{$rel(e_i, e_j)$ in $R_{sorted}$}{
	\uIf{$e_i$ not in $Exec$ \textbf{and} $e_j$ not in $Exec$}
		{
			\tcp{Suppose $score(e_i) \geq score(e_j)$} 
			$S_i \xleftarrow{execute} q_i$; $Exec.add(e_i)$\tcp*[r]{$S_i$:event ID set}
			$S_j \xleftarrow[S_i]{execute} q_j$; $Exec.add(e_j)$\;
			$T \gets S_i \times S_j \mid _{rel(e_i, e_j)}$\tcp*[r]{\textbf{create} tuple set from $S_i$ and $S_j$, then filter by $rel(e_i, e_j)$}
			$M.put(e_i, T)$; $M.put(e_j, T)$\;
		}
    	\uElseIf{Either of $\{e_i, e_j\}$ in $Exec$}
    		{
    			\tcp{Suppose $e_i$ in $Exec$}
    			$S_j \xleftarrow[S_i]{execute} q_j$; $Exec.add(e_j)$\;
    			$T \gets M.get(e_i)$; $T' \gets T\times S_j \mid _{rel(e_i, e_j)}$\tcp*[r]{\textbf{update} tuple set using $S_j$ and $rel(e_i, e_j)$}
    			$replaceVals(M, T, T')$; $M.put(e_j, T')$\;
    		}
    	\Else{
    		$T_i \gets M.get(e_i)$; $T_j \gets M.get(e_j)$\;
    		\uIf{$T_i = T_j$}{
    			$T' \gets T_i\mid _{rel(e_i, e_j)}$\tcp*[r]{\textbf{filter} tuple set}
    			$replaceVals(M, T_i, T')$\;
		}
		\Else{
			$T' \gets T_i\times T_j \mid _{rel(e_i, e_j)}$\tcp*[r]{\textbf{merge} tuple sets}
			$replaceVals(M, T_i, T')$; $replaceVals(M, T_j, T')$\;
    		}
    	}

%	\If{not prefix\_equal($paths[i], paths[i+1], p\text{-}threshold$)}{
%		$breakpoints$.append($i+1$)\;
%	}
}

4. \For{$e_i\in E$ \textbf{and} $e_i$ not in $Exec$}{
	$S_i \xleftarrow{execute} q_i$; $Exec.add(e_i)$; $M.put(e_i, S_i)$\;
}

5. 
\While{$unique(M.values()) > 1$}{
	Pick $T_i$, $T_j$ from $M.values()$, such that $T_i \neq T_j$\;
	$T' \gets T_i\times T_j$\tcp*[r]{\textbf{merge} tuple sets}
	$replaceVals(M, T_i, T')$; $replaceVals(M, T_j, T')$\;
}

6. Return $unique(M.values())$\;
%Return $S_1 \times S_2 \times \cdots S_n \mid _{R} \in T$. 

\vspace{1ex}
% function
\Fn{replaceVals (M, T, T')}{
	Replace all values $T$ stored in $M$ with $T'$\;
%	\For{$key$ in $M.keys()$}{
%		\If{$M.get(key) = T$}{
%			$M.put(key, T')$\;
%		}
%	}
}

\caption{Relationship-based scheduling}\label{alg:schedule}
\end{algorithm}

%%%%%%%%%

Our empirical results (Sec.~\ref{subsec:performance-single} and~\ref{subsec:performance-mpp}) demonstrate that the number of constraints work well in approximating the pruning power of event patterns in a broad set of queries,
even though they may not accurately represent the size of the results returned by event patterns.

%%%%%%
\myparatight{Time Window Partition}
The \dsl query engine leverages temporal properties of the data to further speed up the execution of synthesized data queries:
the engine partitions the time window of a data query into sub-queries with smaller time windows, and executes them in parallel.
Currently, our system splits the time window into days for a query over a multi-day time window.
\section{Deployment and Evaluation}
\label{sec:eval}

We deployed the \dsl system in NEC Labs America comprising 150 hosts (10 servers, 140 employee stations).
We performed a series of attacks based on known exploits in the deployed environment and constructed 46 \dsl queries to investigate these attacks,
demonstrating the expressiveness of \dsl.
To evaluate the effectiveness of \dsl in supporting timely attack investigation, we evaluate the query \emph{efficiency} and \emph{conciseness} against existing systems:
PostgreSQL~\cite{postgresql}, Neo4j~\cite{neo4j}, Splunk~\cite{splunk}.
%As \dsl system can be built on top of PostgreSQL (for small organizations) and Greenplum (for larger enterprises),
We also evaluate the efficiency offered by our data query scheduler (Sec.~\ref{subsec:optimization}) in both storage settings: PostgreSQL and Greenplum.
In total, our evaluations use 857GB of real system monitoring data (16 days; 2.5 billion events).

\subsection{Evaluation Setup}
The evaluations are conducted on a database server with an Intel(R) Xeon(R) CPU E5-2660 (2.20GHz), 64GB RAM,
and a RAID that supports four concurrent reads/writes.
Neo4j databases are configured by importing system entities as nodes and system events as relationships.
Greenplum databases are configured to have 5 segment nodes that can effectively leverage the concurrent reads/writes of RAID. 
For each \dsl query (except anomaly queries), we construct semantically equivalent SQL, Cypher, and Splunk SPL queries.
We measure the execution time and the conciseness of each query.
%, and clean the database cache every time before the execution. 
Note that we omit the performance evaluation of Splunk since the community version is limited to 500MB per day and the enterprise version is prohibitively expensive ($\$$1,900 per GB).
Nevertheless, Splunk's limited support for joins~\cite{splunkjoin} makes it inappropriate for investigating multi-step attack behaviors.
Due to the limited expressiveness of SQL and Cypher, we cannot compare the anomaly queries (\eg Query~\ref{case:stail:anomaly}).
All queries are available on our \emph{project website~\cite{aiql}}.

\eat{
\myparatight{Omitted Approaches}
We omit the evaluation of Neo4j since the case study evaluations already demonstrate that relational databases have better performance than graph databases due to better support for joins in our query tasks.
We omit the evaluation of Splunk since the community version is limited to 500MB per day and the enterprise version is prohibitively expensive ($\$$1,900 per GB). Nevertheless, Splunk's limited support for joins~\cite{splunkjoin} makes it inappropriate for investigating multi-step attack behaviors:
(1) Splunk does not provide indexes to support efficient joins and cartesian product of log entries has to be used for joining two events;
(2) Due to potential performance instability~\cite{splunklimit}, Splunk limits the number of results used for joins to be less than 50,000. 
%
%Nonetheless, we can easily extend \dsl to synthesize Splunk queries,  
%and apply our domain optimizations.
We cannot evaluate against Elasticsearch~\cite{elasticsearch}, a log query tool similar to Splunk, since it does not support joins and cannot directly express event relationships.

}

%%%%%%%%%%%%%%%%
\subsection{Case Study: APT Attack Investigation}
\label{subsec:case}

\begin{figure}[t]
	%\vspace{-0.1cm}
	\centering
	\includegraphics[width=0.45\textwidth]{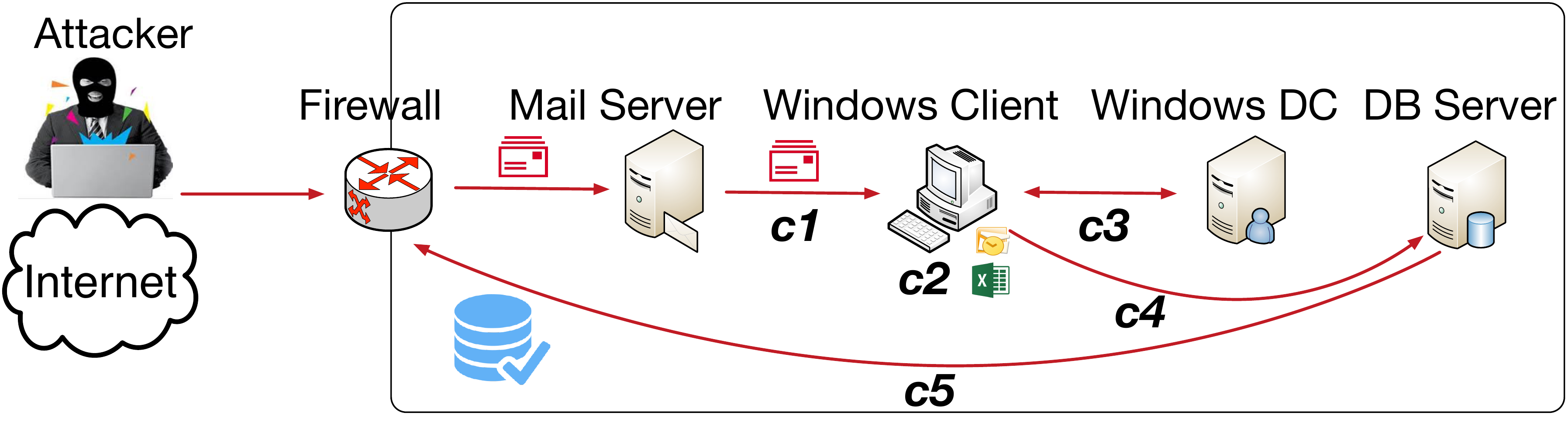}
	%\caption{Diagram for attach reproduction scenario}
	\caption{Environmental setup for the APT attack}
	\label{fig:outlook-apt}
%	\vspace*{-1ex}
\end{figure}

We conduct a case study by asking white hat hackers to perform an APT attack in the deployed environment, as shown in Fig.~\ref{fig:outlook-apt}. 
Below are the attack steps:
\eat{
First, we conduct a case study by performing an APT attack constructed from existing studies~\cite{aptfireeye,aptsymantec} (Fig.~\ref{fig:outlook-apt}). 
To mimic the typical enterprise environment, we deployed a virtual system in our deployed enterprise environment, which consists of a gateway router, a mail server, a Windows client, a Windows domain controller, and a SQL database server. 
To be close to the real APT attack where a portion of hosts are compromised by the attacker, we asked a group of white hat hackers to penetrate into the system using several exploits
and steal the valuable information stored in the database server. 
The successful attack requires the following attack steps:
}

\begin{itemize}[label={\arabic*.}, noitemsep, topsep=1pt, partopsep=1pt, listparindent=\parindent, leftmargin=*]
	\item[\emph{c1}] \emph{Initial Compromise}: The attacker sends a crafted email to the victim. The email contains an Excel file with a malicious macro embedded. 
	
	\item[\emph{c2}] \emph{Malware Infection}: The victim opens the Excel file through the Outlook mail client and runs the macro, which downloads and executes a malware (CVE-2008-0081~\cite{cveexcel}) to open the backdoor to the attacker.
	
	\item[\emph{c3}] \emph{Privilege Escalation}: The attacker enters the victim's machine through the backdoor, scans the network ports to discover the IP address of the database, and runs the database cracking tool ({\tt gsecdump.exe}) to obtain the credentials of the user database.
	
	\item[\emph{c4}] \emph{Penetration into Database Server}: Using the credentials, the attacker penetrates into the database server and delivers a VBScript to drop another malware, which creates another backdoor to the attacker.
	
	\item[\emph{c5}] \emph{Data Exfiltration}: With the access to the database server, the attacker dumps the database content using {\tt osql.exe} and sends the data dump back.
	% to his host.
\end{itemize}

\myparatight{Anomaly Detectors}
We deployed two anomaly detectors based on existing solutions~\cite{anomalysurvey,idsbook,fileanomaly}. 
The first detector is deployed on the database server, which monitors network data transfer and emits alerts when the transfer amount is abnormally large.
The second detector is deployed on the Windows client, which monitors process creation and emits alerts when a process starts an unexpected child process.
These detectors may produce false positives, and we need tools like \dsl to investigate the alerts before taking any further action.

%%%%%%%%%%%%%%%%%%%%%%%%%%%%%%%5

\subsubsection{Attack Investigation Procedure}
\label{case:investigation}

Our investigation assumes no prior knowledge of the detailed attack steps but merely the detector alerts. 
We start with these alerts and iteratively compose \dsl queries to investigate the entire attack sequence. 

%%%%%%%%%%
\myparatight{Step c5}
We first examine the alerts reported by the database server detector, and identify a suspicious external IP ``XXX.129'' (obfuscated for privacy). 
Existing network traffic detectors usually cannot capture the precise process information~\cite{networkids,networkids2}. 
Thus, we first compose an anomaly \dsl query that computes moving average (SMA3) to find processes which transfer a large amount of data to this suspicious IP.

\begin{lstlisting}[captionpos=b, caption={\dsl anomaly query for large file transfer}, label={case:stail:anomaly}]
(at "mm/dd/2017") // date (obfuscated)
agentid = xxx // SQL database server (obfuscated)
window = 1 min, step = 10 sec
proc p write ip i[dstip="XXX.129"] as evt
return p, avg(evt.amount) as amt
group by p
having (amt > 2 * (amt + amt[1] + amt[2]) / 3)
\end{lstlisting}
%\vspace*{-2ex}

Query~\ref{case:stail:anomaly} finishes execution within 4 seconds and identifies a suspicious process ``sbblv.exe''.
%Note that we are unable to easily find such anomalous process using general database queries like SQL or Cypher,
%since they lack support for sliding window and history state comparison.
We then compose a multievent \dsl query to find the data sources for this process (Query~\ref{case:stail:c5:starter}).
%which files were processed by such suspicious process.
\begin{lstlisting}[captionpos=b, caption={Starter \dsl query for \emph{c5}}, label={case:stail:c5:starter}]
(at "mm/dd/2017")
agentid = xx // SQL database server (obfuscated)
proc p1["%sbblv.exe"] read || write file f1 as evt1
proc p1 read || write ip i1[dstip="XXX.129"] as evt2
with evt1 before evt2
return distinct p1, f1, i1, evt1.optype, evt1.access
\end{lstlisting}
%\vspace*{-2ex}

We identify a suspicious file ``BACKUP1.DMP'' for \incode{f1} out of the other normal DLL files. We investigate its creation process and find ``sqlservr.exe", which is a standard SQL server process with verified signature. Thus, we speculate that the attacker penetrates into the SQL server, dumps the data (``BACKUP1.DMP''), and sends the data back to his host (``XXX.129''). We verify this by checking that ``osql.exe" process is started by ``cmd.exe" (OSQL utility is often involved in many SQL database attacks). Query~\ref{case:stail:c5:comp} gives the complete query for investigating the step \emph{c5}.

\begin{lstlisting}[captionpos=b, caption={Complete \dsl query for \emph{c5}}, label={case:stail:c5:comp}]
(at "mm/dd/2017")
agentid = xxx // SQL database server (obfuscated)
proc p1["%cmd.exe"] start proc p2["%osql.exe"] as evt1
proc p3["%sqlservr.exe"] write file f1["%backup1.dmp"] as evt2
proc p4["%sbblv.exe"] read file f1 as evt3
proc p4 read || write ip i1[dstip="XXX.129"] as evt4
with evt1 before evt2, evt2 before evt3, evt3 before evt4
return distinct p1, p2, p3, f1, p4, i1
\end{lstlisting}
\myparatight{Steps c4-c1}
The investigation for \emph{c4}-\emph{c1} is similar to \emph{c5}, including iterative query execution and editing.
% We include the investigation details in Appendix~\ref{appendix:case}.
%and we omit the discussion and the involved queries due to space limitations.
%We omit the discussion of the investigation process and the involved queries for \emph{c4} - \emph{c1} due to space limitations. 
In total, we constructed 26 multievent queries and 1 anomaly query to successfully investigate the APT attack, touching 119GB of data/422 million events.
% entire APT attack sequence.
%The investigation details are available on our project website~\cite{aiql}.

\eat{
In order to penetrate into the database server, the attacker needs to obtain the DB server IP address and DB administrator credential, which are typically stored in the Windows domain controller. We write a query to confirm that a connection is made from the Windows domain controller to the database server. 
Since there are many ways for escalating privilege, without any detection alerts, we are not able to investigate further.
For c1, receiving email is a normal behavior at the Windows Client and thus we compose a query only to confirm whether any email from the IP of the attacker host sends an email to the email server.
%The investigation and queries for these two steps are in Appendix~\ref{appendix:case:c3c1}.
The involved queries can be found on our anonymous website~\cite{stail}.
}

%%%%%%%%%%%%%%%%%%%%%%%%%%%%%%%%%%%%%
%%%%%%%%%%%%%%%%%%%%%%%%%%%%%%%%%%%%%
\subsubsection{Evaluation Results}
\label{case:eval-results}

As we can see, attack investigation is an iterative process that revises queries:
(1) latter iterations add more event patterns based on the selected results from the former queries,
and (2) 4-5 iterations are needed before finding a complete query with 5-7 event patterns.
Thus, \emph{slow response} and \emph{verbose specification} could greatly impede the effectiveness and efficiency of the investigation.

\begin{figure}[!tp]
%\vspace{-0.1cm}
\centering
\includegraphics[width=0.48\textwidth]{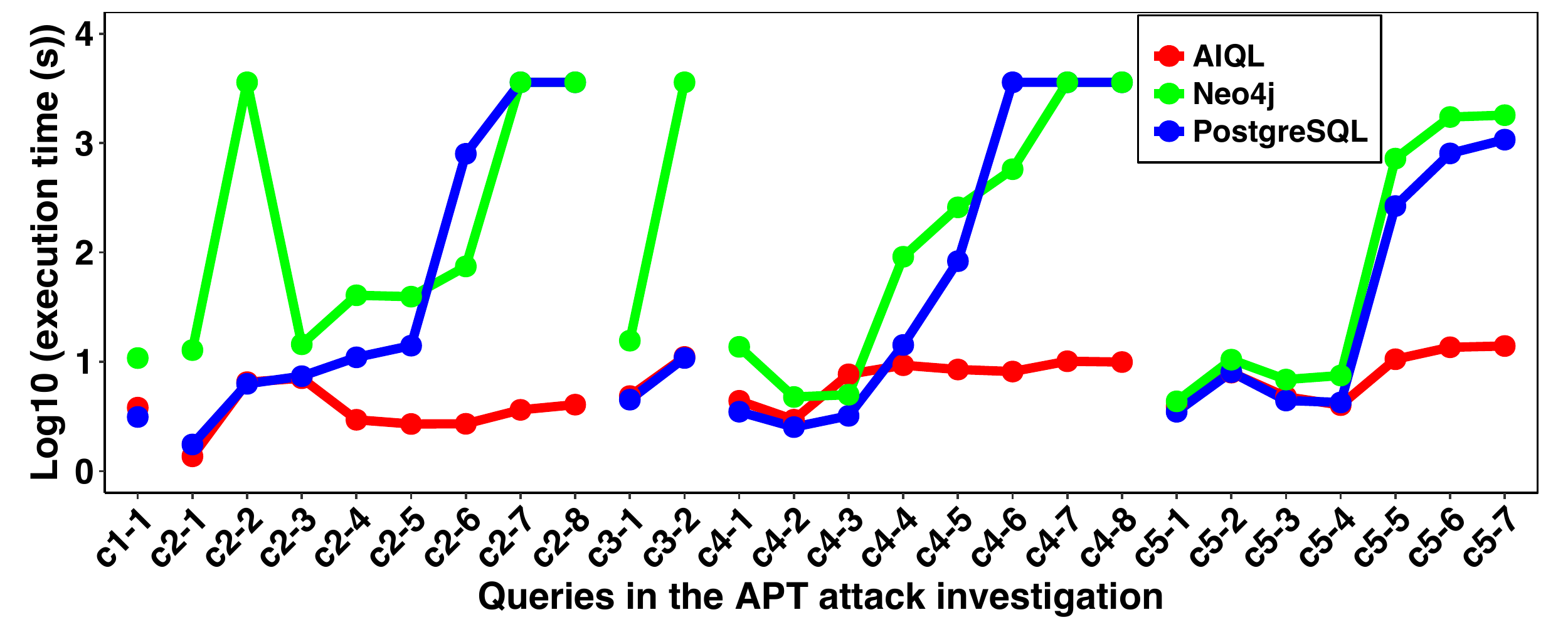}
\vspace{-2ex}
\caption{Log10-transformed query execution time}
%Log10-transformed execution time of queries on the \dsl system, PostgreSQL and Neo4j. We plot the points at 3600s for queries that cannot finish within 1 hour.}
\label{fig:case-study-results}
\end{figure}

\myparatight{End-to-End Execution Efficiency}
Fig.~\ref{fig:case-study-results} shows the execution time of \dsl queries, 
%built on top of PostgreSQL, 
SQL queries in PostgreSQL, and Cypher queries in Neo4j. 
For evaluation fairness, PostgreSQL and Neo4j databases store the same copies of data and employ the same schema and index designs as \dsl, but they do not employ our domain-specific data storage optimizations such as spatial and temporal partitioning, nor our scheduling optimizations.\footnote{Fine-grained evaluations of the \dsl scheduling are in Sec.~\ref{subsec:scheduling-eval}.}
Table~\ref{tab:case} shows aggregate statistics for investigating each attack step, including the number of queries, the number of event patterns, and the total investigation time (second).
We observe that:
(1) Neo4j generally runs slower than PostgreSQL, due to the lack of support for efficient joins;
% in graph databases;
(2) PostgreSQL and Neo4j become very slow when the query becomes complex and the number of event patterns (hence the required table joins) becomes large. Many large queries in PostgreSQL and Neo4j cannot finish within 1 hour (e.g., c2-7, c2-8, c4-7, c4-8);
(3) all \dsl queries finish within 15 seconds, and the performance of the queries grows linearly with the number of event patterns (rather than the exponential growth in PostgreSQL and Neo4j),
demonstrating the effectiveness of our domain-specific storage optimizations and query scheduling.
(4) the total investigation time is $\sim$5.9 hours for PostgreSQL and $\sim$7.5 hours for Neo4j, 
which is a significant bottleneck for a timely attack investigation. 
%which significantly prevents security analysts from obtaining the timely feedback on queries for iteratively revising. 
In contrast, the total investigation time for \dsl is within 3 minutes (124x speedup over PostgreSQL and 157x speedup over Neo4j).

\eat{
Note that our measurement was done by cleaning database cache before every execution.
% for fair comparison. 
In the real-life investigation without cleaning cache, the \dsl system can achieve further performance speedup by leveraging the data locality, since adjacent investigation queries often preserve certain level of semantic similarity (\eg the latter query is constructed from the former query by adding another event pattern).
Due to the ineffective scheduling on many constraints, SQL queries for large number of patterns bring too much data into memory and cache, 
flushing the cache and benefiting very little from the results of previous queries.}

\myparatight{Conciseness}
The largest \dsl query is c4-8 with 7 event patterns, 25 query constraints, 109 words, and 463 characters (excluding spaces). The corresponding SQL query contains 77 constraints (3.1x larger), 432 words (4.0x larger), and 2792 characters (6.0x larger). The corresponding Cypher query contains 63 constraints (2.5x larger), 361 words (3.3x larger), and 2570 characters (5.6x larger).
As the attack behaviors become more complex, SQL and Cypher queries become verbose with many joins and constraints, 
posing challenges for constructing the queries for timely attack investigation.
% stats for c4-8
% num chars: 463, 2792, 2570
% num words: 109, 431, 361
% num of constraints: 25, 77, 63

% adopt an iterative procedure for query tuning and execution.

\eat{
\begin{table}[t]
	\centering
	\caption{Case study statistics}\label{tab:case}
	\begin{adjustbox}{width=0.3\textwidth}
		%	\scriptsize
		\begin{tabular}{|r|r|r|r|r|r|}
			\hline
			Step &Queries &Event Patterns	& \dsl (s)	& SQL (s)	& Cypher (s)\\\hline
			c1 	&1	&3		&3.803	&3.127				& 10.813	\\\hline
			c2 	&8	&27		&30.9815	&8038.681				& 10981.688\\\hline
			c3 	&2	&4		&15.9195	&15.33 				& 3615.593	\\\hline
			c4 	&8	&35		&61.006	&10906.708			& 8150.55	 \\\hline
			c5 	&7	&18		&58.7505	&2166.475 			& 4285.375	\\\hline
			All 	&26	&87		&170.4605	&21130.321 			& 27044.019	\\\hline
		\end{tabular}
	\end{adjustbox}
\end{table}
}

\begin{table}[!tp]
	\centering
	\caption{Aggregate statistics for case study}\label{tab:case}
	\begin{adjustbox}{width=0.44\textwidth}
		%	\scriptsize
		\begin{tabular}{|l|l|l|l|l|l|}
			\hline
			Attack Step &$\#$ of Queries 	& $\#$  of Evt Patterns	& \dsl (s)	& PostgreSQL (s)	& Neo4j (s)	\\\hline
			c1 	&1		&3			&3.8			&3.1				& 10.8		\\\hline
			c2 	&8		&27			&31.0		&8038.7			& 10981.7		\\\hline
			c3 	&2		&4			&15.9		&15.3 			& 3615.6		\\\hline
			c4 	&8		&35			&61.0		&10906.7			& 8150.6	 	\\\hline
			c5 	&7		&18			&58.8		&2166.5 			& 4285.4		\\\hline
			All 	&26		&87			&170.5		&21130.3 			& 27044.1		\\\hline
		\end{tabular}
	\end{adjustbox}
\end{table}

%%%%%%%%%%%%%%%%

\begin{figure*}
	\center
	\begin{subfigure}[H]{0.23\textwidth}
		\includegraphics[width=\linewidth]{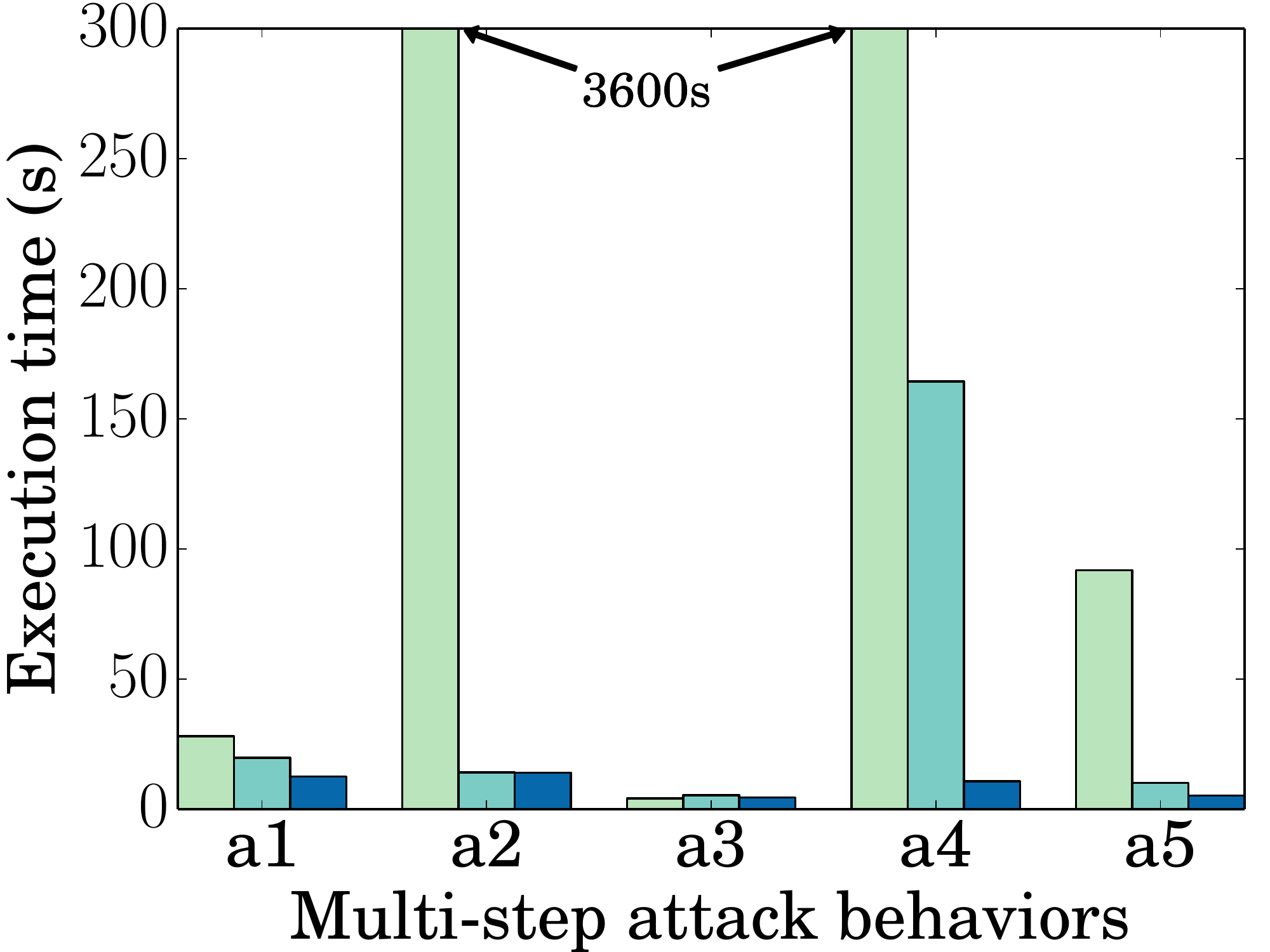}
		%\caption{}
		\label{fig:eval-att}
	\end{subfigure}%
	\hfill
	\begin{subfigure}[H]{0.23\textwidth}
		\includegraphics[width=\linewidth]{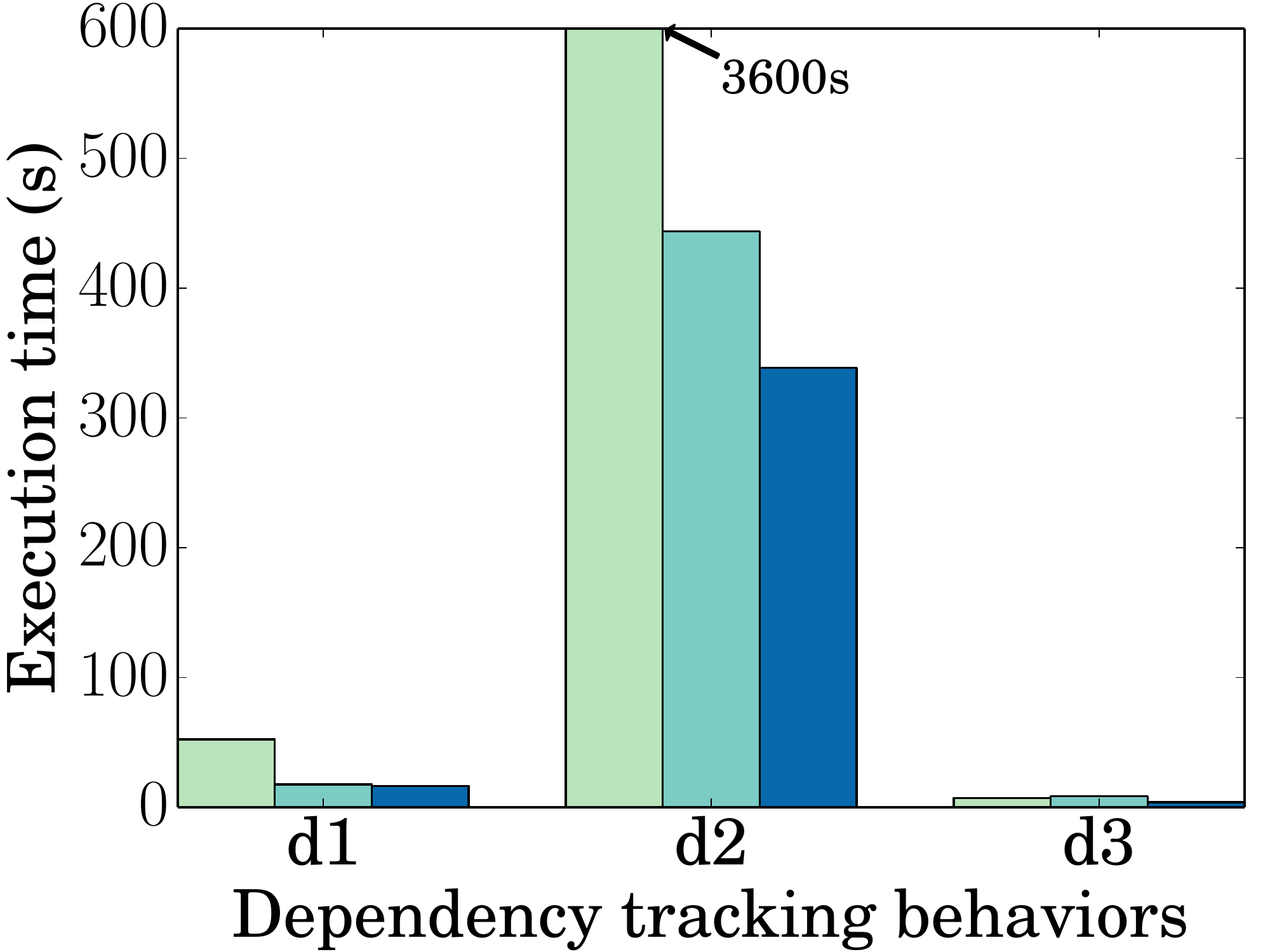}
		%\caption{}
		\label{fig:eval-dep}
	\end{subfigure}
	\hfill	
	\begin{subfigure}[H]{0.23\textwidth}
%		\vspace*{-3ex}
		\includegraphics[width=\linewidth]{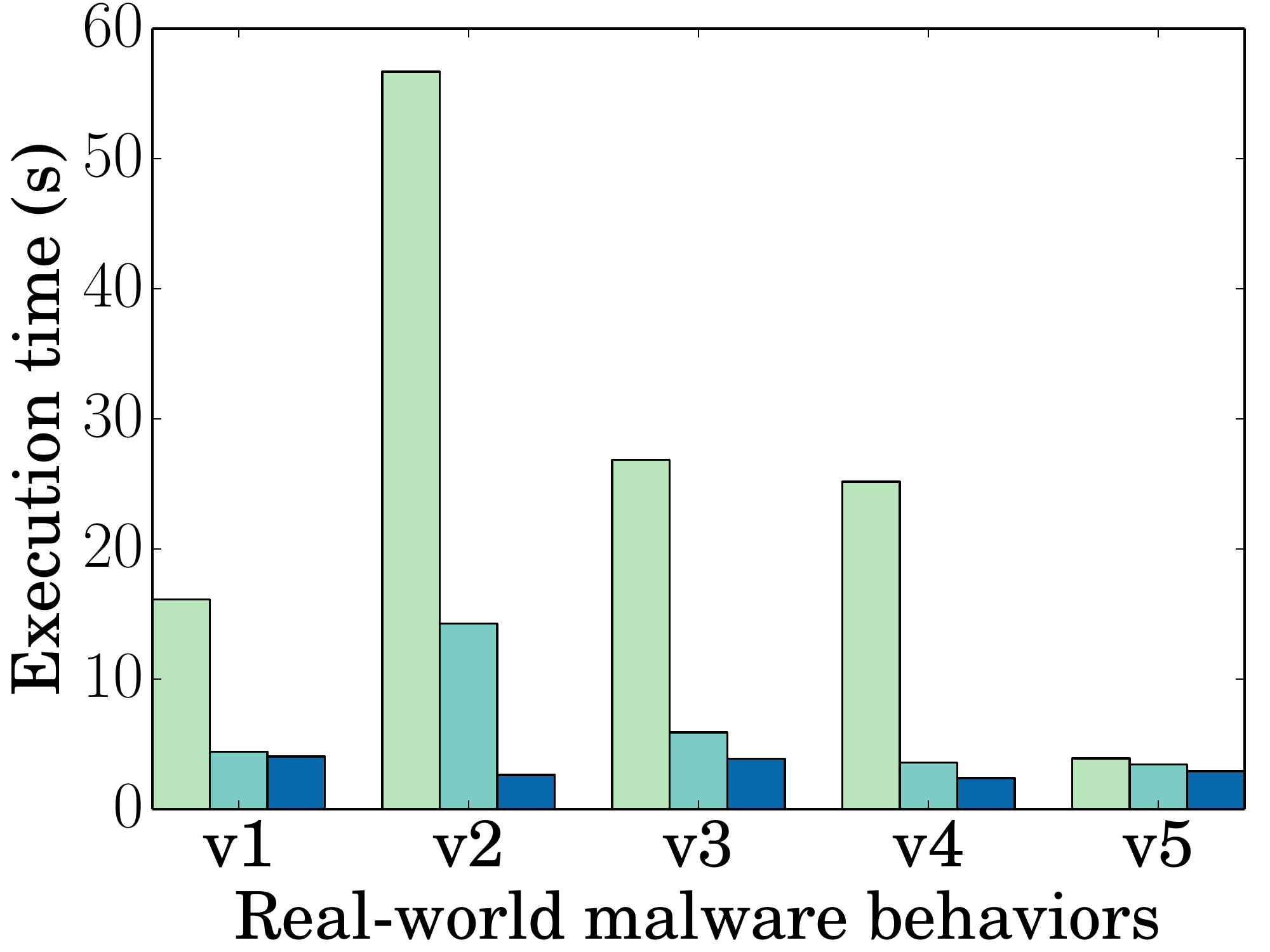}
		%\caption{}
		\label{fig:eval-vir}
	\end{subfigure}%
	\hfill
	\begin{subfigure}[H]{0.23\textwidth}
%		\vspace*{-3ex}
		\includegraphics[width=\linewidth]{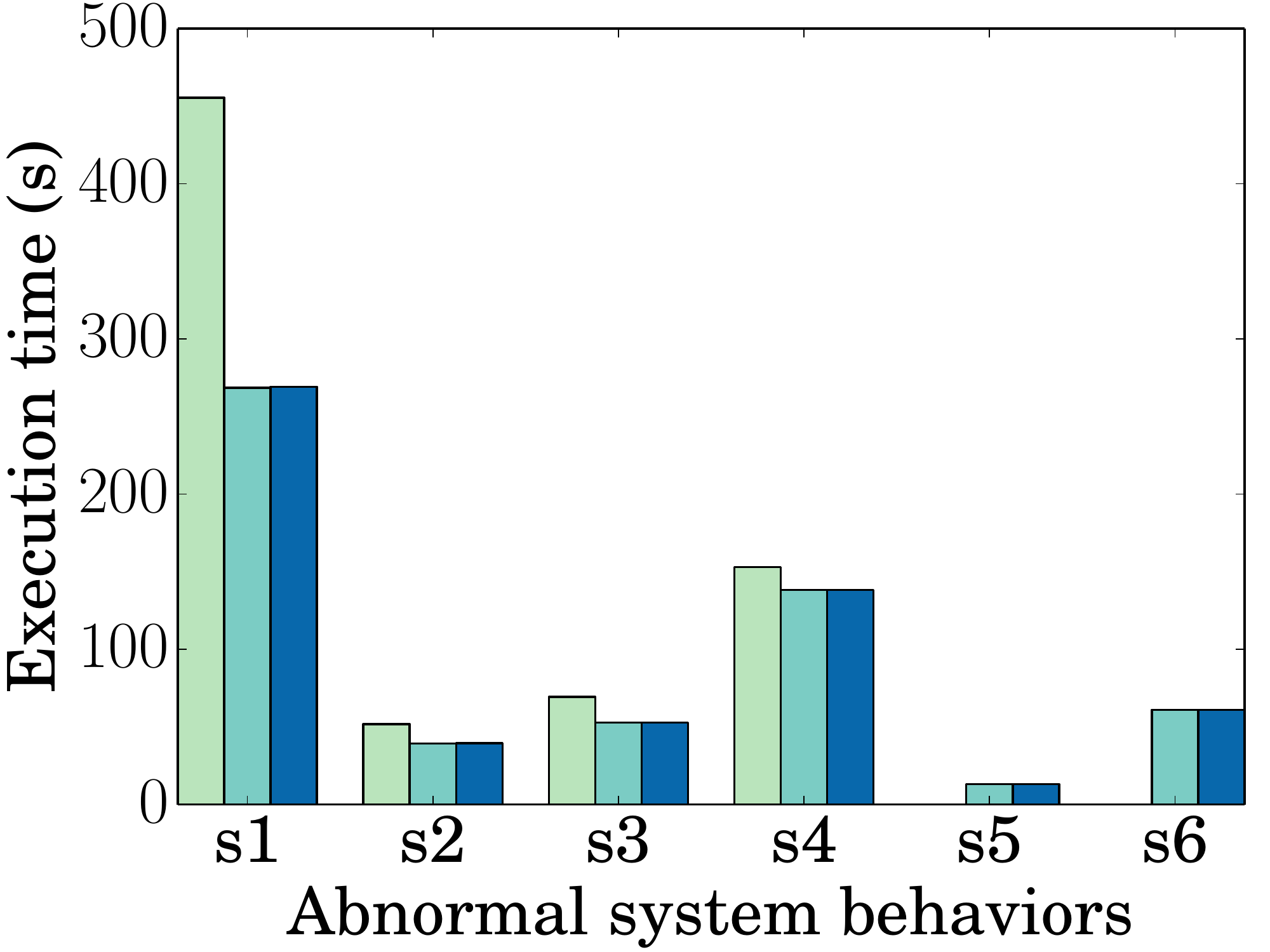}
		%\caption{}
		\label{fig:eval-sus}
	\end{subfigure}
	\vspace*{-1.5ex}

	\begin{subfigure}[H]{0.1\textwidth}
	%		\vspace*{-3ex}
			\includegraphics[width=\linewidth]{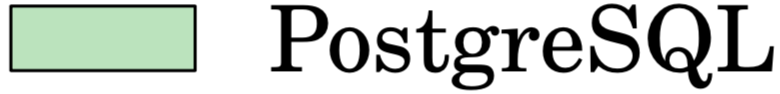}
			%\caption{}
	%		\label{fig:eval-sus}
	\end{subfigure}%
	\hspace{3cm}
	\begin{subfigure}[H]{0.1\textwidth}
%		\vspace*{-3ex}
		\includegraphics[width=\linewidth]{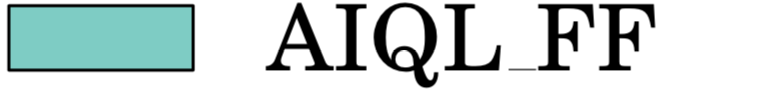}
		%\caption{}
%		\label{fig:eval-sus}
	\end{subfigure}
	\hspace{3cm}
	\begin{subfigure}[H]{0.1\textwidth}
%		\vspace*{-3ex}
		\includegraphics[width=\linewidth]{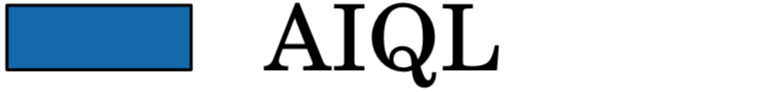}
		%\caption{}
%		\label{fig:eval-sus}
	\end{subfigure}
	
	\caption{
		Query execution time of the scheduling employed by PostgreSQL, \dslff, and \dsl (single-node)
	}
	\label{fig:eval-schedule-single}
	%\vspace*{-4ex}
\end{figure*}

\subsection{Performance Evaluation}
\label{subsec:scheduling-eval}
%We build our \dsl system on top of .
% for larger organizations.
% (leveraging distributed storage and process-level parallelism to further scale up the system).
We evaluate the performance of \dsl in both storage settings (PostgreSQL and Greenplum) by constructing 19 \dsl queries for a broad set of attack behaviors, touching 738GB/2.1 billion events. 
Particularly, we are interested in the efficiency speedup provided by the \dsl scheduling (Sec.~\ref{subsec:optimization}) in comparison with PostgreSQL scheduling and Greenplum scheduling.

%%%%%%%%%%%%%%%%%%%%%%
\subsubsection{Attack Behaviors}
\label{subsubsec:riskyeval}

\myparatight{Multi-Step Attack Behaviors}
We asked white hat hackers to launch another APT attack using different exploits (details available on~\cite{aiql}). We then constructed 5 \dsl queries for investigating the attack steps (\emph{a1-a5}).

%We asked white hat hackers to launch another APT attack using different exploits (details available on~\cite{aiql}). We adopt a similar investigation procedure as Sec.~\ref{case:investigation} and constructed \dsl queries for investigating the 5 attack steps (\emph{a1-a5}).

\myparatight{Dependency Tracking Behaviors}
We performed causal dependency tracking of origins of Chrome update executables (\emph{d1}) and Java update executables (\emph{d2}). 
We performed forward dependency tracking of the ramification malware \incode{info_stealer}
% to analyze its impact 
(\emph{d3}).

\myparatight{Real-World Malware Behaviors}
%We executed 5 real-world malware samples in our deployed environment.
We obtained a dataset of free malware samples from VirusSign~\cite{virussign}.
%, which contains the malware samples obtained from 11/15/15 to 04/23/16. 
%Each malware sample is accompanied with a behavior report generated by the Automated Malware Analysis System (VSAMAS)~\cite{virussign}, which describes its malware categorization and behaviors.
%In this dataset, we 
We then
randomly selected 5 malware samples (Table~\ref{tab:malware}) from the 3 largest categories: \emph{Autorun}, \emph{Sysbot}, and \emph{Hooker}.
%These samples have complex behaviors, such as creating a suspicious file and reading from a suspicious IP address.
We executed the 5 selected samples in the deployed environment and constructed \dsl queries by analyzing the accompanied behavior reports~\cite{virussign} (\emph{v1-v5}).

\eat{\emph{Virus.Autorun} and \emph{Trojan.Sysbot} are the two largest categories, which contain 24.28\% and 21.25\% of malware samples, respectively.
	For these two largest categories, we randomly selected two malware samples, and executed the malware samples in a Windows VM that is installed with our monitoring agent to collect the system monitoring data.
	We used Process Monitor from Windows Sysinternals~\cite{xx} to examine the system activities associated with the malware, in order to confirm that the malware is able to run successfully.
	We selected three other malware samples from  \emph{Trojan.Hooker} and \emph{Virus.Sysbot} categories.}

\begin{table}[t]
	\centering
	\caption{Selected malware samples from Virussign}\label{tab:malware}
	\begin{adjustbox}{width=0.44\textwidth}
		\begin{tabular}{|l|l|l|}
			\hline
			ID	& Name		& Category\\\hline
			v1	&7dd95111e9e100b6243ca96b9b322120	&Trojan.Sysbot\\\hline
			v2	&425327783e88bb6492753849bc43b7a0	&Trojan.Hooker\\\hline
			v3	&ee111901739531d6963ab1ee3ecaf280	&Virus.Autorun \\\hline
			v4 	&4e720458c357310da684018f4a254dd0	&Virus.Sysbot \\\hline
			v5	&7dd95111e9e100b6243ca96b9b322120	&Trojan.Hooker \\\hline
		\end{tabular}
	\end{adjustbox}
	%	\vspace*{-1ex}
	
	%	\vspace*{-4ex}
\end{table}

%%%%%%%%%%%%%%%%%%

%%%%%%%%%%%%%
\myparatight{Abnormal System Behaviors}
We evaluated 6 abnormal system behaviors based on security experts' knowledge:
(1) \emph{s1}: command history probing; 
%(2) \emph{s2}: processes except Apache listening to port 80; 
(2) \emph{s2}: suspicious web service; 
(3) \emph{s3}: frequent network access;  
(4) \emph{s4}: erasing traces from system files;
%and frequency-based abnormal behaviors
(5) \emph{s5}: network access spike;
(6) \emph{s6}: abnormal file access.
Note that 
for \emph{s5} and \emph{s6}, we did not construct SQL, Cypher, or Splunk queries, due to their lack of support for sliding window and history state comparison.

\subsubsection{Efficiency in PostgreSQL}
\label{subsec:performance-single}

%We evaluate the efficiency offered by the \dsl execution scheduling (Sec.~\ref{subsec:optimization}) in single-node databases. 
We select two baselines:
(1) PostgreSQL databases that \emph{employ our data storage optimizations (Sec.~\ref{subsec:datastorage})}. 
Note that this setting is different from the end-to-end efficiency evaluation in Sec.~\ref{case:eval-results}, because here we want to rule out the speedup offered by the data storage component;
(2) \dsl with fetch-and-filter scheduling (denoted as \dslff; Sec.~\ref{subsec:optimization}). 
We measure the execution time of the 19 queries in Sec.~\ref{subsubsec:riskyeval}.
%( including 1,816,419,047 file events, 90,475,871 process events, and 155,937,724 network events) 

\myparatight{Evaluation Results}
Fig.~\ref{fig:eval-schedule-single} shows the execution time of queries in PostgreSQL, \dslff, and \dsl.
We observe that:
(1) the scheduling employed by PostgreSQL is inefficient in executing complex queries. In particular, PostgreSQL cannot finish executing \emph{a2}, \emph{a4}, and \emph{d2} within 1 hour;
(2) the scheduling employed by \dslff and \dsl is more efficient than PostgreSQL, with 19x and 40x speedup, respectively;
(3) the relationship-based scheduling employed by \dsl is more efficient than the fetch-and-filter scheduling employed by \dslff.
% (2.4x speedup). 
  
%($speed\text{-}up = \frac{execution\_time\_SQL}{execution\_time\_\dsl}$).

%%%%%%%%%%%%%%%%

\begin{figure*}
	\center
	\begin{subfigure}[H]{0.23\textwidth}
		\includegraphics[width=\linewidth]{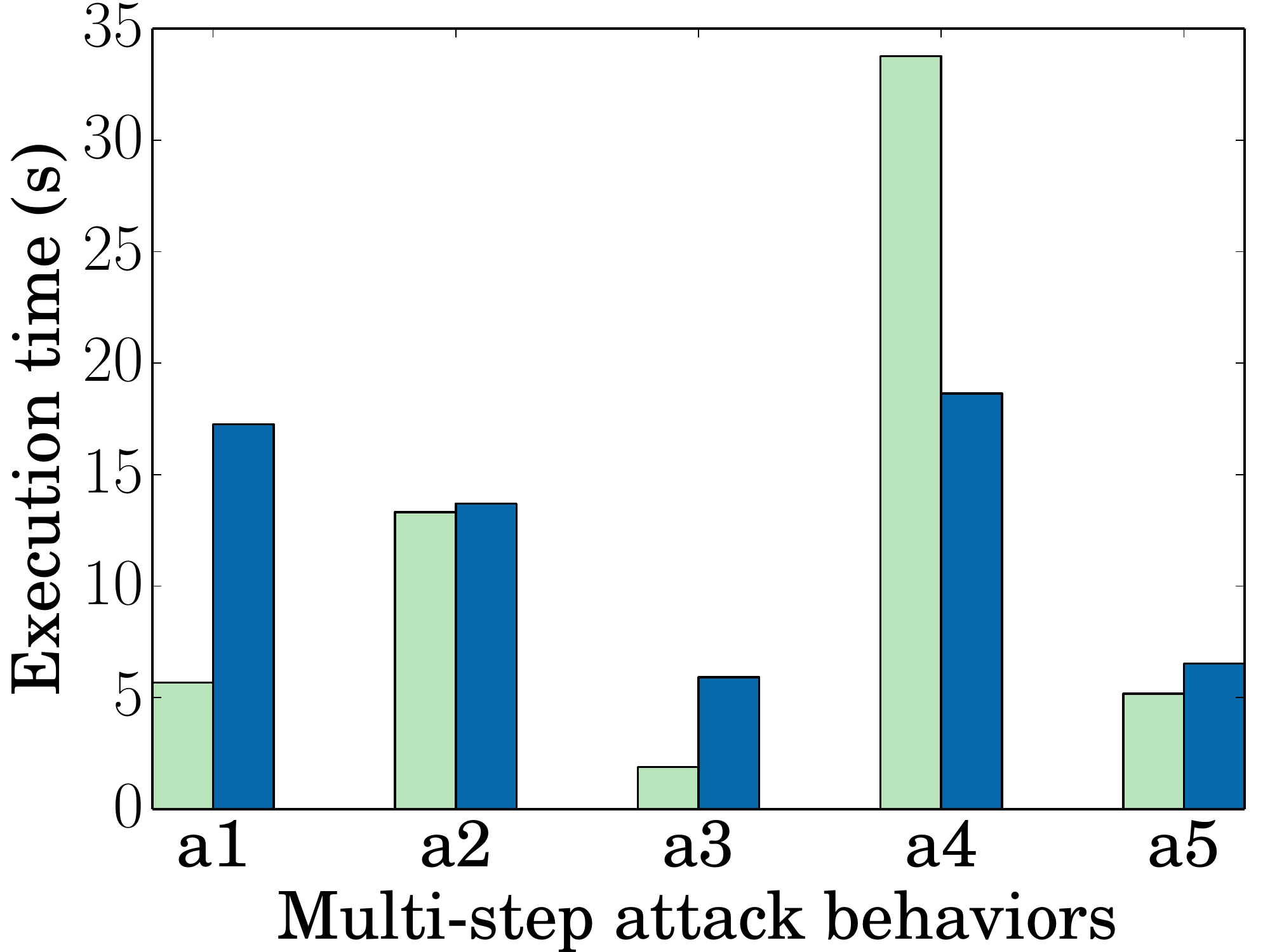}
		%\caption{}
		\label{fig:eval-att}
	\end{subfigure}%
	\hfill
	\begin{subfigure}[H]{0.23\textwidth}
		\includegraphics[width=\linewidth]{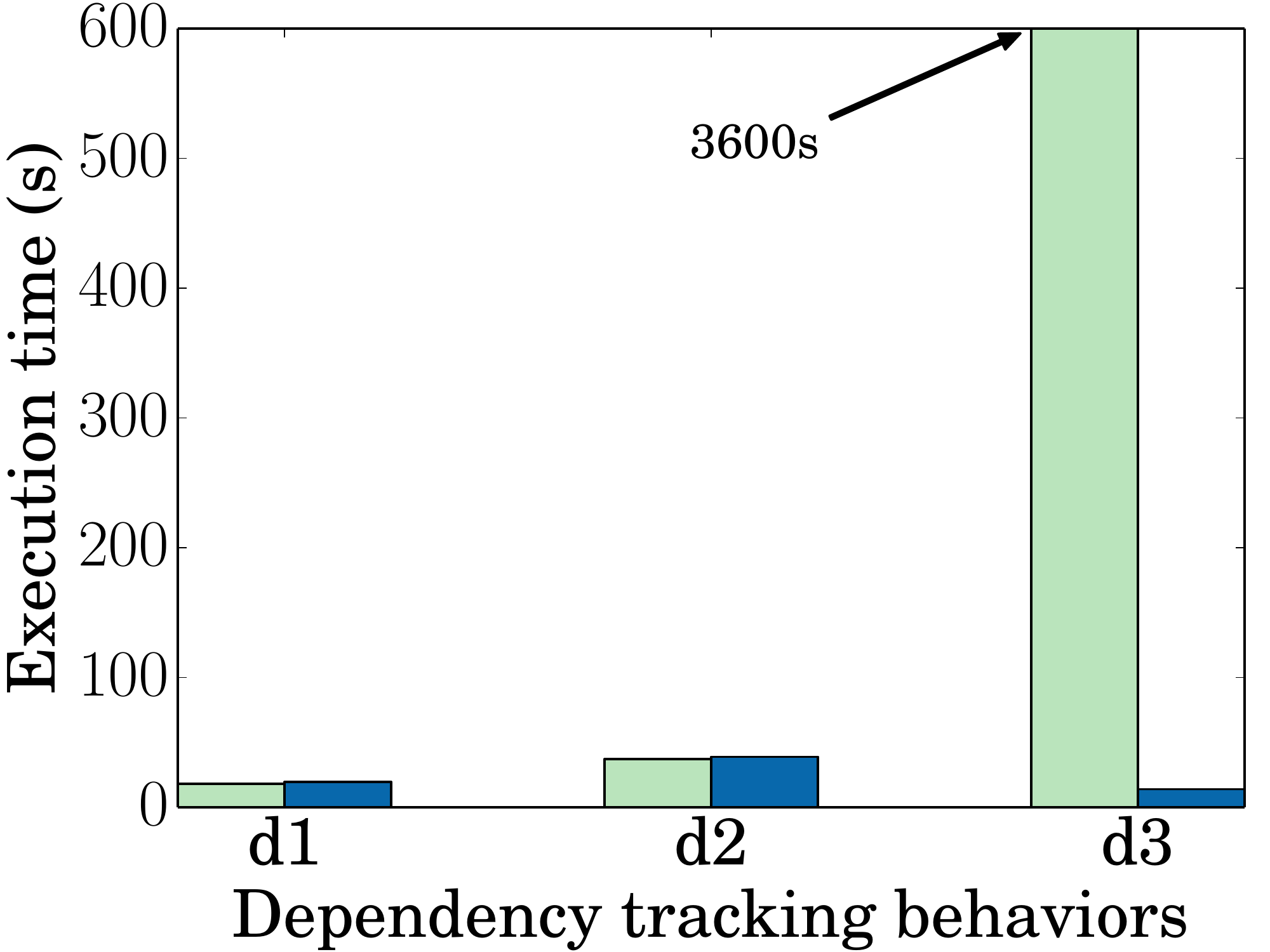}
		%\caption{}
		\label{fig:eval-dep}
	\end{subfigure}
	\hfill	
	\begin{subfigure}[H]{0.23\textwidth}
%		\vspace*{-3ex}
		\includegraphics[width=\linewidth]{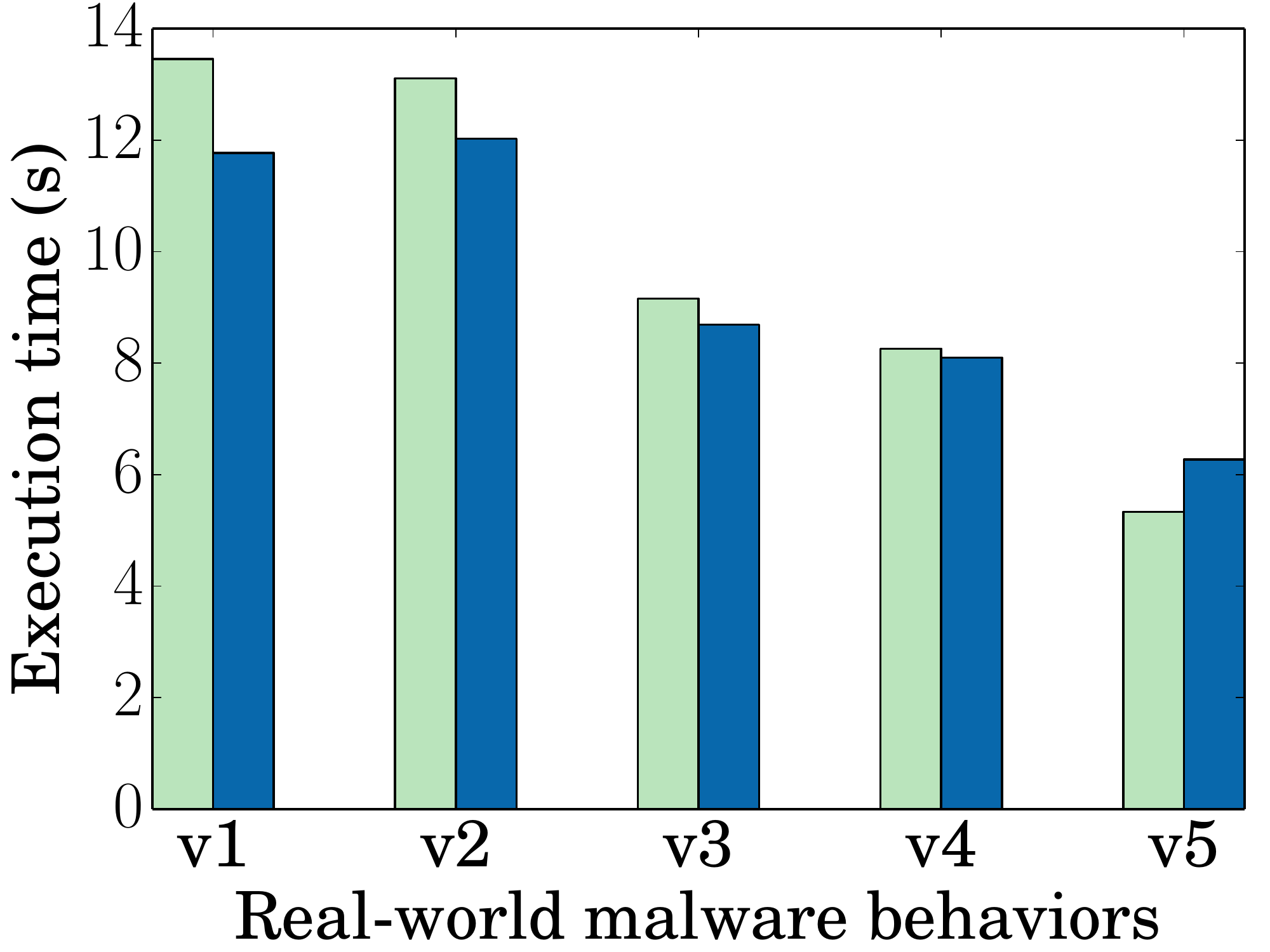}
		%\caption{}
		\label{fig:eval-vir}
	\end{subfigure}%
	\hfill
	\begin{subfigure}[H]{0.23\textwidth}
%		\vspace*{-3ex}
		\includegraphics[width=\linewidth]{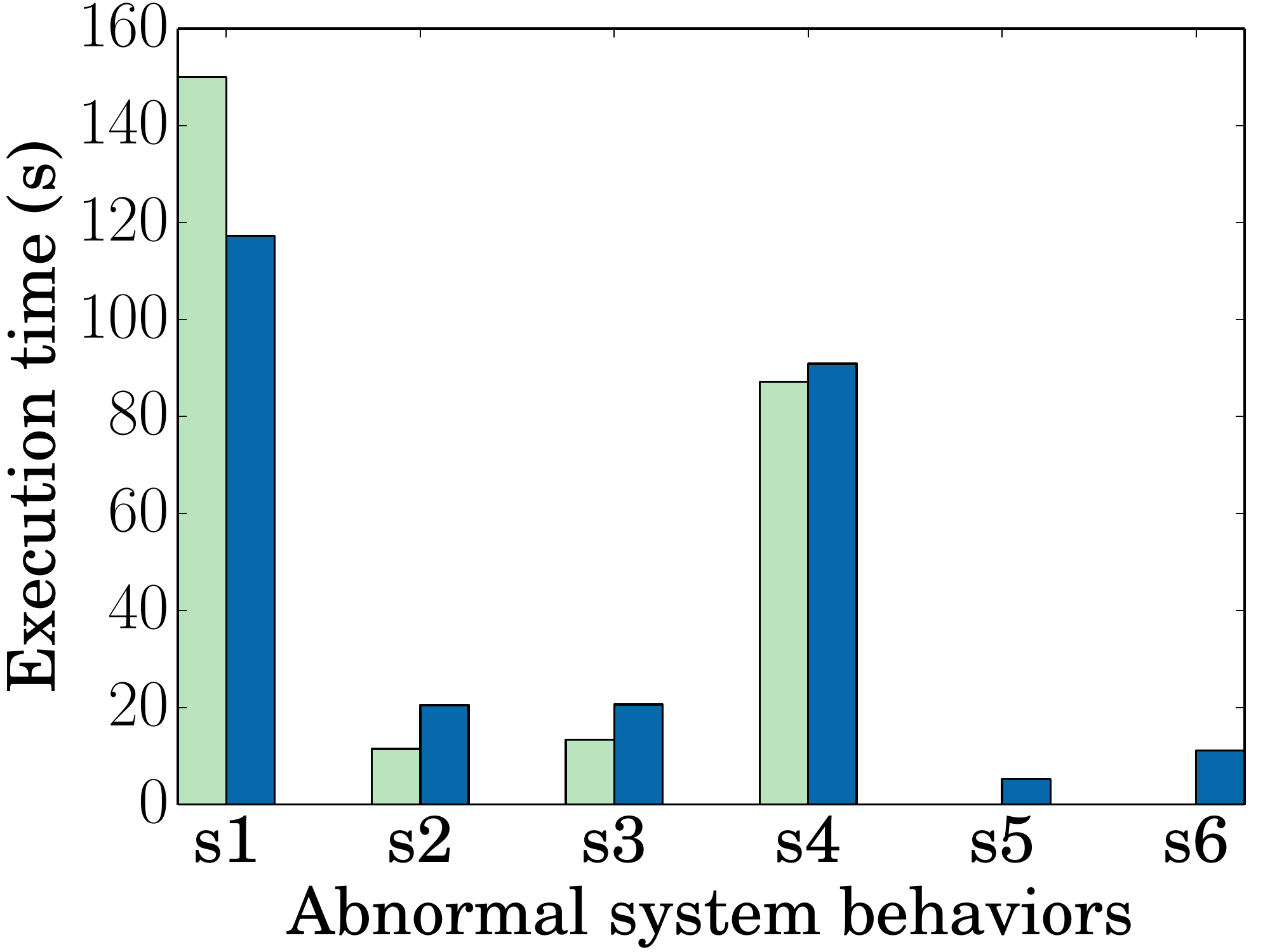}
		%\caption{}
		\label{fig:eval-sus}
	\end{subfigure}
	\vspace*{-2ex}
	
	\begin{subfigure}[H]{0.1\textwidth}
	%		\vspace*{-3ex}
			\includegraphics[width=\linewidth]{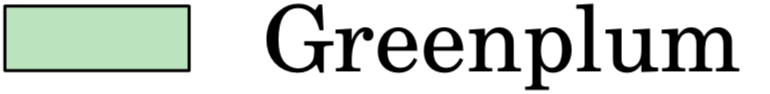}
			%\caption{}
	%		\label{fig:eval-sus}
	\end{subfigure}%
	\hspace{3cm}
	\begin{subfigure}[H]{0.1\textwidth}
%		\vspace*{-3ex}
		\includegraphics[width=\linewidth]{l3.png}
		%\caption{}
%		\label{fig:eval-sus}
	\end{subfigure}
	
	\caption{
		Query execution time of the scheduling employed by Greenplum and \dsl (parallel)
	}
	\label{fig:eval-schedule-mpp}
%	\vspace*{-1ex}
\end{figure*}

\subsubsection{Efficiency in Parallel Databases}
\label{subsec:performance-mpp}
We compare the performance of \dsl scheduling in the Greenplum storage with the Greenplum scheduling (i.e., running SQLs).
%with Greenplum without our scheduling (i.e., running SQLs)
%using the 19 queries in Sec.~\ref{subsubsec:riskyeval}. 
As in Sec.~\ref{subsec:performance-single}, the Greenplum databases also \emph{employ our data storage optimizations}. 
%%Similar to Sec~\ref{subsec:performance-single}, we
%We select Greenplum databases that employ our data storage optimizations as a baseline.
% We omit the evaluation of \dslff.

\myparatight{Evaluation Results} 
Fig.~\ref{fig:eval-schedule-mpp} shows the execution time of queries in Greenplum and \dsl.
We observe that:
(1) in most cases, our scheduling in parallel settings achieves a comparable performance as Greenplum scheduling;
(2) in certain cases (\eg \emph{a4}, \emph{d3}), our scheduling is significantly more efficient than Greenplum scheduling;
(3) the average speedup over Greenplum is 16x.
The results show that without our semantics-aware model, Greenplum distributes the storage of events based on their incoming orders (which is arbitrary).
%Thus, events coming from a host are not evenly distributed and parallel search on certain hosts cannot be properly optimized.
On the contrary, our data model allows Greenplum to evenly distribute events in a host, and achieves more efficient parallel search.

%%%%%%%%%%%%%%%%

\begin{figure*}[!tp]
	%\vspace{-0.1cm}
	\centering
	%\vspace*{-0.5ex}
	\begin{subfigure}[H]{0.31\textwidth}
		\includegraphics[width=\linewidth]{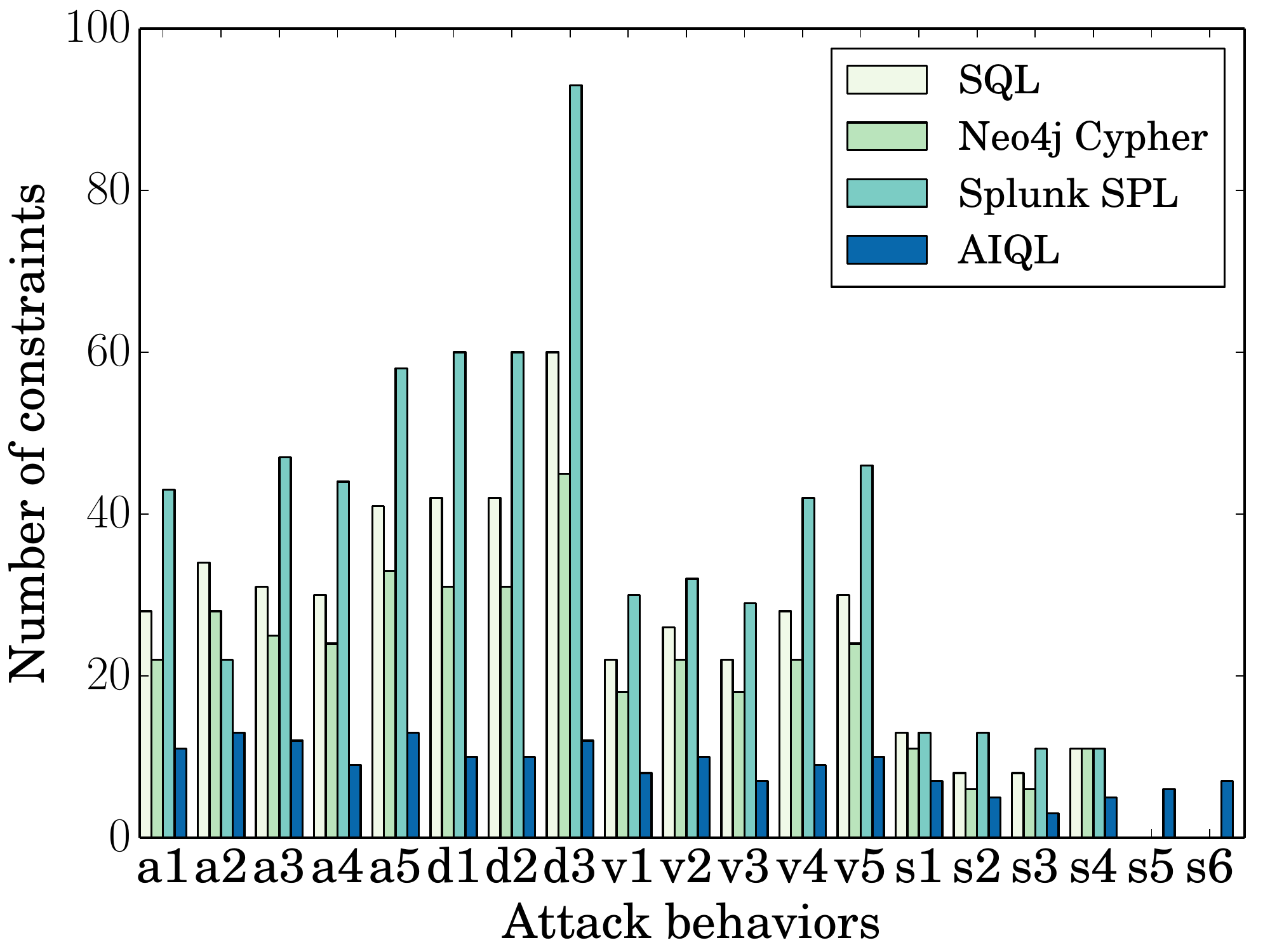}
		\caption{Number of constraints}
		\label{fig:conciseness:constraint}
	\end{subfigure}%
	%	\hspace{1cm}
	\begin{subfigure}[H]{0.31\textwidth}
		\includegraphics[width=\linewidth]{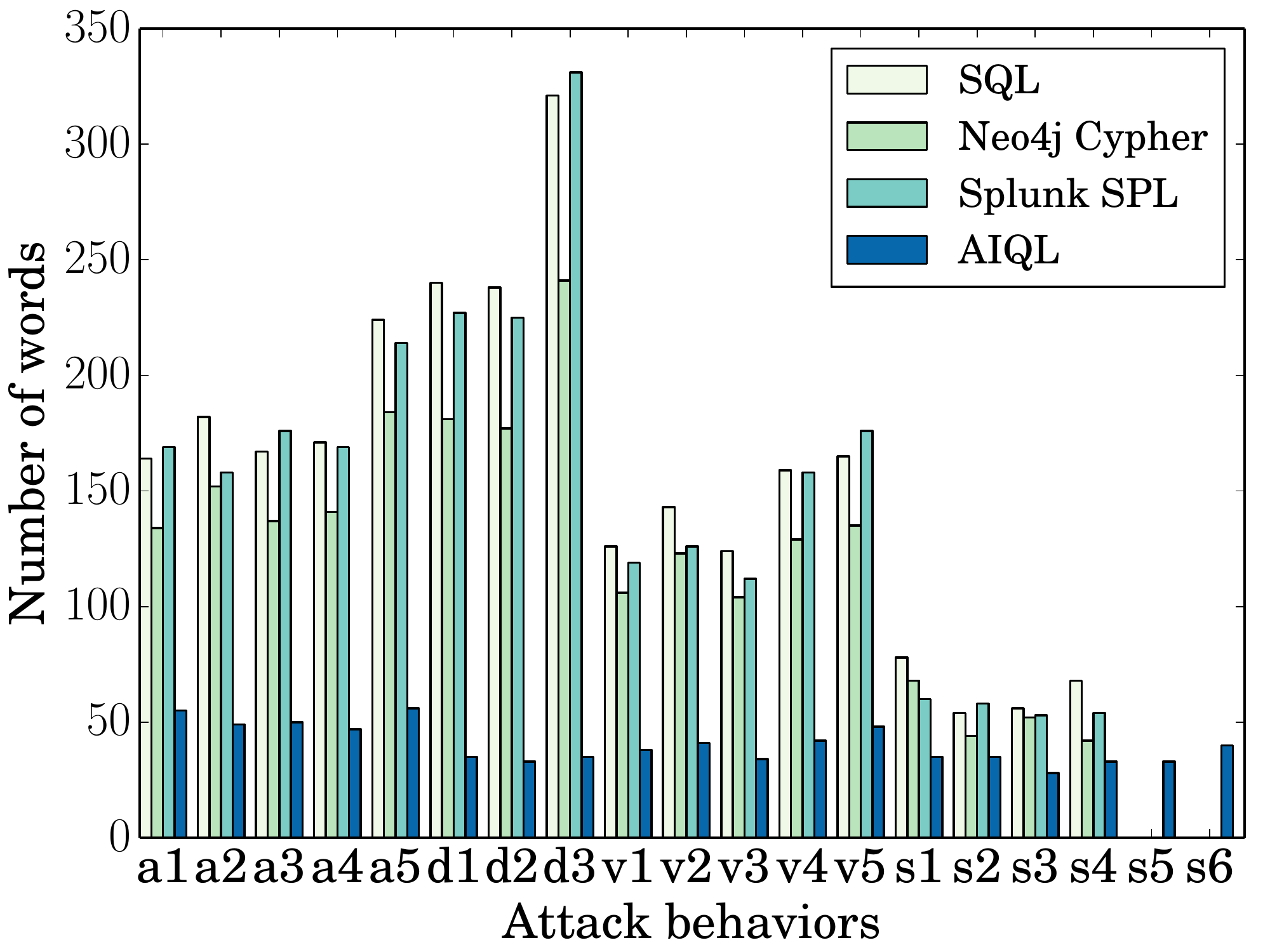}
		\caption{Number of words}
		\label{fig:conciseness:word}
	\end{subfigure}%
	%	\hspace{1cm}
	\begin{subfigure}[H]{0.31\textwidth}
		\includegraphics[width=\linewidth]{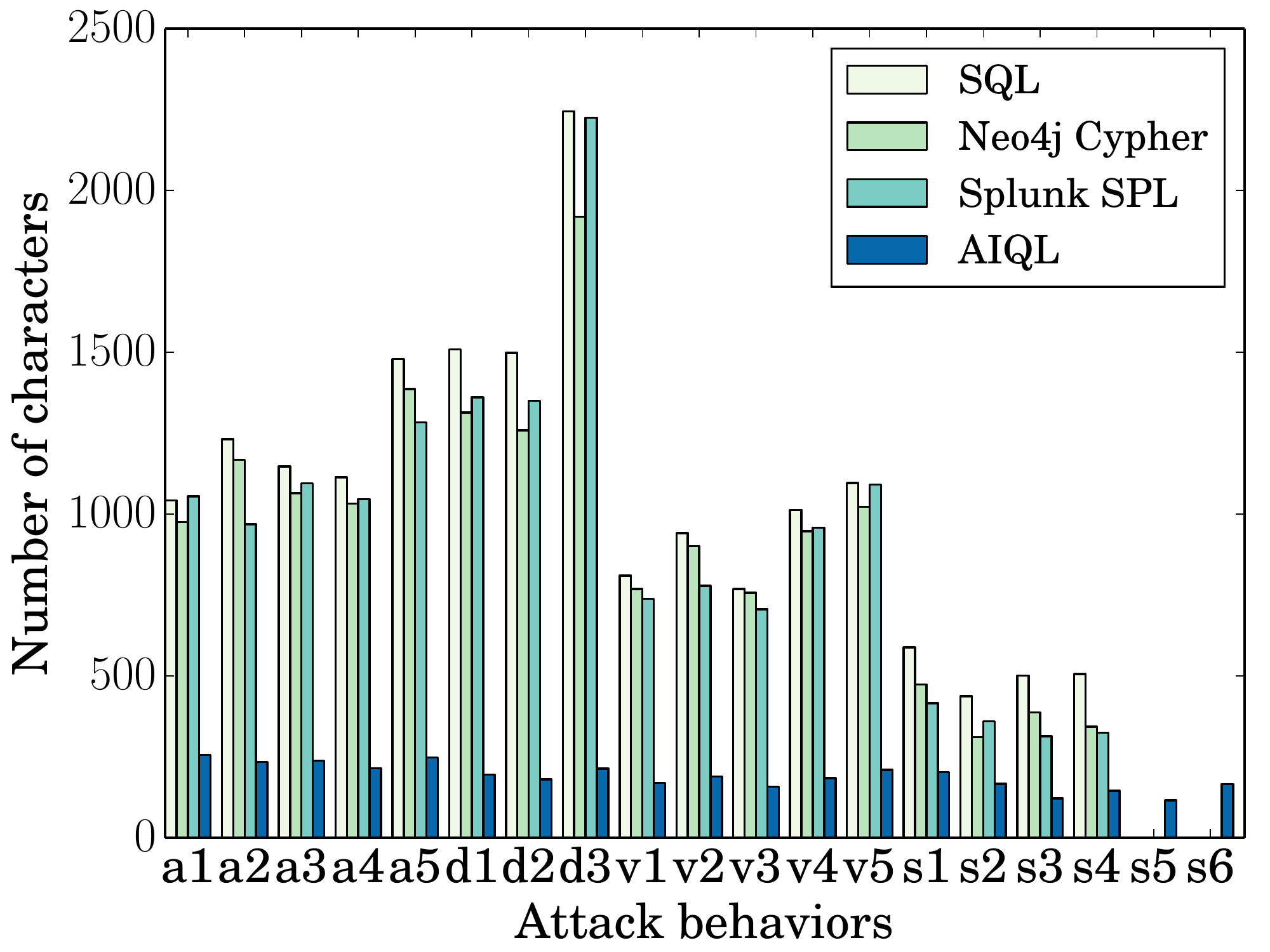}
		\caption{Number of characters}
		\label{fig:conciseness:char}
	\end{subfigure}%

	\caption{Conciseness evaluation of queries written in \dsl, SQL, Neo4j Cypher, and Splunk SPL
		%Number of characters of \dsl, SQL, and Splunk queries for \emph{a1-a5}, \emph{d1-d3}, \emph{v1-v5}, and \emph{s1-s6}, excluding spaces and comments. For \emph{s5} and \emph{s6}, we do not evaluate SQL and Splunk queries since they do not support sliding time windows and the access to history states.
	}
	\label{fig:conciseness}
%	\vspace*{-1ex}
\end{figure*}

%%%%%%%%%%%%%%%%

\subsection{Conciseness Evaluation}
\label{subsec:conciseness}

%We conduct conciseness evaluations by performing four major types of attacks in the deployed environment, and constructed 19 \dsl queries.
%We conduct conciseness evaluations using the \dsl queries in Sec.~\ref{subsubsec:riskyeval}.
We evaluate the conciseness of queries that express the 19 attack behaviors in Sec.~\ref{subsubsec:riskyeval} in three metrics: the number of query constraints, the number of words, and the number of characters (excluding spaces).

\begin{table}[t]
	\centering
	%	\caption{Conciseness improvement of \dsl queries over SQL, Neo4j Cypher, and Splunk SPL queries}\label{tab:conciseness}
	\caption{Conciseness improvement statistics}\label{tab:conciseness}
	\begin{adjustbox}{width=0.44\textwidth}
		\begin{tabular}{|l|l|l|l|}
			\hline
			Metrics		& \dsl/SQL	& \dsl/ Cypher	& \dsl/Splunk SPL\\\hline
			$\#$ of constraints		& 3.0x		& 2.4x 	& 4.2x\\\hline
			$\#$ of words 		& 3.9x 		& 3.1x 	& 3.8x\\\hline
			$\#$ of characters		& 5.3x		& 4.7x 	& 4.7x\\\hline
		\end{tabular}
	\end{adjustbox}
	%	\vspace*{-4ex}
\end{table}

%%%%%%%%%%%%%%%%%%%%%%%%%%%%%%%%%%%%
\myparatight{Evaluation Results}
%\label{subsubsec:conciseness-eval-results}
Fig.~\ref{fig:conciseness} shows the conciseness metrics of \dsl, SQL, Neo4j Cypher, and Splunk SPL queries. Table~\ref{tab:conciseness} shows the average improvement of \dsl queries over other queries. 
We observe that \emph{\dsl is the most concise query language} in terms of all three metrics and all attack behaviors:
SQL, Neo4j Cypher, and Splunk SPL contain at least 2.4x more constraints, 3.1x more words, and 4.7x more characters than \dsl.
In contrast to SQL, Cypher, and SPL which employ lots of joins on tables or nodes, \dsl provides high-level constructs for spatial/temporal constraints, relationship specifications, constraints chaining, and context-aware syntax shortcuts, making the queries much more concise.

\section{Discussion}

\myparatight{Query Scheduler}
Our data query scheduler estimates the pruning score of an event pattern based on its number of constraints.
This can be improved by (1) considering the number of records in different hosts and different time periods and (2) constructing a statistical model of constraint pruning power.
Additionally, the query scheduler may partition the time window uniformly based on the data volume.
%, so that the data volume in each split window is roughly the same.
Such strategies require further analysis of the domain data statistics to infer the proper data volume for splitting,
% and the overhead for sharing the data, 
which we leave for future work.

\myparatight{System Entities and Data Reduction}
%Besides files, processes, and network connections in our current data model, in 
In the future work, we plan to add registry entries in Windows and pipes in Linux to expand the monitoring scope. We also plan to incorporate more finer granularity system monitoring, such as execution partition~\cite{mpi,protracer}
% to record more precise activities of processes~\cite{execpartition,execpartition2} 
and in-memory data manipulations~\cite{panda,r2}. To handle the increase of data size, we plan to explore more aggressive data reduction techniques in addition to existing solutions~\cite{loggc,reduction} to make the system more scalable.

%\dsl can be easily extensible to support these new entities by expanding the entities and operations in the data model.

%\myparatight{Real-Time Query}
%\dsl enables attack investigation over historical results, allowing security analysts to recover attack sequences by correlating multiple system activities.
%We plan to extend \dsl with the ability to query real-time streams of system monitoring data, assisting real-time anomaly analysis.

%%%%%%%%%%%%%%%%%%%%%%%%%%%%%%%%%%%%%%%%%%%%%%
\section{Related Work}
\label{sec:related}

\myparatight{Security-Related Languages}
There also exist domain-specific languages in a variety of security fields that have a well-established corpus of low level
algorithms, such as threat descriptions~\cite{cybox,taxii,stix}, 
%cryptographic systems~\cite{crypto1,crypto2,oblivm}, 
secure overlay networks~\cite{mace,networklang}, and network intrusions~\cite{chimera,lambda,Sommer:2014:HAE:2663716.2663735,Vallentin:2016:VUP:2930611.2930634}.
% and obfuscations~\cite{Marionette:2015}.
These languages 
%are explicitly designed to solve domain-specific problems, 
%providing 
provide specialized constructs for their particular problem domain.
% and eschewing irrelevant features.
In contrast to these languages, 
the novelty of \dsl
% is its focus on assisting attack investigations, 
focuses on querying attack behaviors,
including 
(a) providing specialized constructs for system interaction patterns/relationships and 
abnormal behaviors;
(b) optimizing query execution over system monitoring data.
Splunk~\cite{splunk} and Elasticsearch~\cite{elasticsearch} are distributed search and analytics engine for application logs,
which provide search languages based on keywords and shell-like piping.
However, these systems lack efficient supports for joins and their languages cannot express abnormal behaviors with history states as \dsl.
Furthermore, our \dsl can be used to investigate the real-time anomalies detected on the stream of system monitoring data,
complementing the stream-based anomaly detection systems~\cite{saql} for better defense.

\eat{
is a platform that automatically parses application logs,
and provides a Unix shell-like Search Processing Language to filter entries of logs based on a combination of keywords.
%It includes solutions for security, compliance, and fraud~\cite{splunksecurity}.
However, the shell-like language is verbose for expressing event patterns and their relationships,
and certainly cannot express abnormal behaviors with history states as \dsl.
Also, Splunk does not provide indexes to support efficient joins, which is critical for the performance in searching multi-step software behaviors. 
Similar to Splunk, Elasticsearch~\cite{elasticsearch} is a distributed search and analytics engine for logs.
But Elasticsearch does not support join of log entries and cannot be used to query event relationships.}

\myparatight{Database Query Languages}
%Database query languages are designed for managing data held in various types of databases.
%Theses languages cannot express all three major types of risky system behaviors.
Relational databases based on SQL~\cite{postgresql,sql} and SPARQL~\cite{sparql} provide language constructs for joins, 
facilitating the specification of relationships among events, but these languages lack constructs for easily chaining constraints among relations (\ie tables).
Graph databases~\cite{neo4j} provide language constructs for chaining constraints among nodes in graphs,
but these databases lack efficient support for joins.
Similarly, NoSQL tools~\cite{chodorow2013mongodb} lack efficient supports for joins.
% and are not suitable for such querying tasks.
%
%Besides existing databases, there also exist query languages in research works for spatio-temporal databases~\cite{spatiotemporaldb}, but the languages are optimized for specifying system behaviors.
Temporal expressions are also introduced to databases~\cite{tquel},
and various time-oriented applications are explored~\cite{timesql}.
Currently, \dsl focuses on the set of temporal expressions that are frequently used in expressing attack behaviors,
which is a subset of the temporal expressions proposed in~\cite{tquel}.
More importantly, none of these languages provide constructs to express frequency-based behavioral models with historical results.

\eat{
There also exist domain-specific languages for specific security applications.
Chimera~\cite{chimera} is a declarative query language for network traffic processing,
which adds structured data types, first-class functions, and dynamic window boundaries
to extend SQL languages to better handle network traffic.
ObliVM~\cite{oblivm} provides a programming framework for secure computation.
It provides a programming language equipped with programming abstractions 
that are user-friendly for non-expert programmers to build oblivious programs.
These languages focus on }

\myparatight{System Defense Based on Behavioral Analytics}
Existing malware detection has looked at various ways to 
build behavioral models to capture malware,
such as sequences of system calls~\cite{staticanalyzer}, 
system call patterns based on data flow dependencies~\cite{malwaresystemcall},
and interactions between benign programs and the operating system~\cite{accessminer}.
Behavioral analytics have also shown promising results for network intrusion~\cite{trafficaggregation,networkmalware} and internal threat detection~\cite{insider}.
%Bhatkar et al.~\cite{dataflowanomaly} proposed a formal notation to represent dataflow properties in terms of several system calls,
%and provided a technique to learn temporal properties involving the arguments of different system calls to capture the flow of security-sensitive data.
%A recent work on behavioral analytics also proposed an active-learning system, AI$^2$~\cite{ai2},
%which collects security analyst' feedback about outlier events and learns supervised models to predict attacks.
These works learn models to detect anomaly or predict attacks, but they do not provide mechanisms for users to perform attack investigation.
Our \dsl system fills such gap by allowing security analysts to query historical events for investigating the reported anomalies.

\section{Conclusion}
\label{sec-conclusions}
We have presented a novel system for collecting attack provenance using system monitoring and assisting timely attack investigation. 
Our system provides (1) domain-specific data model and storage for scaling the storage and the search of system monitoring data,
(2) a domain-specific query language, \emph{Attack Investigation Query Language (\dsl)} that integrates critical primitives for attack investigation,
and (3) an optimized query engine based on the characteristics of the data and the queries to better schedule the query execution.
%We deployed the \dsl system in an anonymous enterprise and conducted a series of evaluations on the query conciseness and execution efficiency. 
Compared with existing 
%database 
systems, our \dsl system greatly reduces the cycle time for iterative and interactive attack investigation. 
% in iteratively revising queries based on previous results for attack investigation.
%In future work, we plan to extend \dsl with the ability to query real-time streams of system monitoring data, assisting real-time enterprise system monitoring.

%%%%%%%%%%%%%%%%%%%%%%%%%%%%%%%%%%%%%%%%%%%%%%

\myparatight{Acknowledgement}
%We would like to thank the anonymous reviewers for their feedback in finalizing this paper. 
This work was partially supported by the National Science Foundation under grants CNS-1553437 and CNS-1409415, Microsoft Research Asia, Jiangsu ``Shuangchuang'' Talents Program, CCF-NSFOCUS ``Kunpeng'' Research Fund, and Alipay Research Fund.
Any opinions, findings, and conclusions
made in this material are those of the authors and do not
necessarily reflect the views of the funding agencies.

%%%%%%%%%%%%%%%%%%%%%%%%%%%%%%%%%%%%%%%%%%%%%%

%\bibliographystyle{ACM-Reference-Format}
%\bibliographystyle{IEEEtranS}
%\bibliography{ref}

{\footnotesize \bibliographystyle{acm}
\bibliography{ref}}

\begin{thebibliography}{10}

\bibitem{aiql}
{AIQL}: Enabling efficient attack investigation from system monitoring data.
\newblock https://sites.google.com/site/aiqlsystem/.

\bibitem{antlr}
{ANTLR}.
\newblock http://www.antlr.org/.

\bibitem{flink}
{Apache Flink}.
\newblock https://flink.apache.org/.

\bibitem{cveexcel}
{CVE-2008-0081}.
\newblock http://www.cve.mitre.org/cgi-bin/cvename.cgi?name=CVE-2008-0081.

\bibitem{cve1}
{CVE-2010-2075}.
\newblock https://cve.mitre.org/cgi-bin/cvename.cgi?name=CVE-2010-2075.

\bibitem{cybox}
{Cyber Observable eXpression (CybOX}$^{TM}$).
\newblock https://cyboxproject.github.io/.

\bibitem{cypher}
{Cypher Query Language}.
\newblock http://neo4j.com/developer/cypher/.

\bibitem{sql-tuning}
Database performance tuning guide.
\newblock https://docs.oracle.com/cd/B19306_01/server.102/b14211/.

\bibitem{ebay}
{eBay Inc. To Ask eBay Users To Change Passwords}.
\newblock http://blog.ebay.com/ebay-inc-ask-ebay-users-change-passwords/.

\bibitem{elasticsearch}
{Elasticsearch}.
\newblock https://www.elastic.co/.

\bibitem{equifax}
The {Equifax} data breach.
\newblock https://www.ftc.gov/equifax-data-breach.

\bibitem{esper}
Esper.
\newblock http://www.espertech.com/products/esper.php.

\bibitem{etw}
{ETW events in the common language runtime}.
\newblock https://msdn.microsoft.com/en-us/library/ff357719(v=vs.110).aspx.

\bibitem{greenplum}
Greenplum.
\newblock http://greenplum.org/.

\bibitem{homedepot}
{Home Depot} confirms data breach at {U.S., Canadian} stores.
\newblock
  http://www.npr.org/2014/09/09/347007380/home-depot-confirms-data-breach-at-u-s-canadian-stores.

\bibitem{neo4j}
Neo4j: The world's leading graph database.
\newblock http://neo4j.com/.

\bibitem{ntp}
{Network Time Protocol} (version 3) specification, implementation and analysis.
\newblock https://tools.ietf.org/html/rfc1305.

\bibitem{opm}
{OPM} government data breach impacted 21.5 million.
\newblock
  http://www.cnn.com/2015/07/09/politics/office-of-personnel-management-data-breach-20-million.

\bibitem{postgresql}
{PostgreSQL}.
\newblock http://www.postgresql.org/.

\bibitem{adware}
Protecting against potentially unwanted programs.
\newblock https://portal.mcafee.com/documents/Show/2096.

\bibitem{siddhi}
{Siddhi}.
\newblock https://github.com/wso2/siddhi.

\bibitem{sparql}
{SPARQL}.
\newblock https://www.w3.org/TR/rdf-sparql-query/.

\bibitem{splunk}
Splunk.
\newblock http://www.splunk.com/.

\bibitem{splunkjoin}
Splunk: joining two searches with common field.
\newblock
  https://answers.splunk.com/answers/105469/joining-two-searches-with-common-field.html.

\bibitem{sql}
{SQL}.
\newblock http://www.iso.org/iso/catalogue_detail.htm?csnumber=45498.

\bibitem{stix}
{Structured Threat Information eXpression (STIX}$^{TM}$).
\newblock http://stixproject.github.io/.

\bibitem{target}
{Target data breach incident}.
\newblock
  http://www.nytimes.com/2014/02/27/business/target-reports-on-fourth-quarter-earnings.html?_r=1.

\bibitem{auditd}
{The Linux audit framework}.
\newblock https://github.com/linux-audit/.

\bibitem{netspike}
Top 5 causes of sudden network spikes.
\newblock
  https://www.paessler.com/press/pressreleases/top_5_causes_of_sudden_spikes_in_traffic.

\bibitem{tc}
Transparent computing.
\newblock http://www.darpa.mil/program/transparent-computing.

\bibitem{taxii}
{Trusted Automated eXchange of Indicator Information (TAXII}$^{TM}$).
\newblock https://taxiiproject.github.io/.

\bibitem{trustwave}
Trustwave global security report 2015.
\newblock
  https://www2.trustwave.com/rs/815-RFM-693/images/2015\_TrustwaveGlobalSecurityReport.pdf.

\bibitem{virussign}
Virussign.
\newblock http://www.virussign.com/.

\bibitem{trustkernel}
{\sc Bates, A., Tian, D., Butler, K. R.~B., and Moyer, T.}
\newblock Trustworthy whole-system provenance for the linux kernel.
\newblock In {\em {USENIX} Security\/} (2015).

\bibitem{chimera}
{\sc Borders, K., Springer, J., and Burnside, M.}
\newblock Chimera: A declarative language for streaming network traffic
  analysis.
\newblock In {\em USENIX Security\/} (2012).

\bibitem{anomalysurvey}
{\sc Chandola, V., Banerjee, A., and Kumar, V.}
\newblock Anomaly detection: A survey.
\newblock {\em CSUR 41}, 3 (2009), 15:1--15:58.

\bibitem{taserdb}
{\sc Chandra, R., Kim, T., Shah, M., Narula, N., and Zeldovich, N.}
\newblock Intrusion recovery for database-backed web applications.
\newblock In {\em SOSP\/} (2011).

\bibitem{chodorow2013mongodb}
{\sc Chodorow, K.}
\newblock {\em MongoDB: The Definitive Guide: Powerful and Scalable Data
  Storage}.
\newblock O'Reilly Media, Inc., 2013.

\bibitem{lambda}
{\sc Cuppens, F., and Ortalo, R.}
\newblock Lambda: A language to model a database for detection of attacks.
\newblock In {\em RAID\/} (2000).

\bibitem{panda}
{\sc Dolan-Gavitt, B., Hodosh, J., Hulin, P., Leek, T., and Whelan, R.}
\newblock Repeatable reverse engineering with panda.
\newblock In {\em PPREW\/} (2015).

\bibitem{saql}
{\sc Gao, P., Xiao, X., Li, D., Li, Z., Jee, K., Wu, Z., Kim, C.~H., Kulkarni,
  S.~R., and Mittal, P.}
\newblock {SAQL}: A stream-based query system for real-time abnormal system
  behavior detection.
\newblock In {\em {USENIX} Security\/} (2018).

\bibitem{taser}
{\sc Goel, A., Po, K., Farhadi, K., Li, Z., and de~Lara, E.}
\newblock The taser intrusion recovery system.
\newblock In {\em SOSP\/} (2005).

\bibitem{r2}
{\sc Guo, Z., Wang, X., Tang, J., Liu, X., Xu, Z., Wu, M., Kaashoek, M.~F., and
  Zhang, Z.}
\newblock R2: An application-level kernel for record and replay.
\newblock In {\em OSDI\/} (2008).

\bibitem{hamilton1994time}
{\sc Hamilton, J.~D.}
\newblock {\em Time series analysis}, vol.~2.
\newblock Princeton University Press, 1994.

\bibitem{wormlog}
{\sc Jiang, X., Walters, A., Xu, D., Spafford, E.~H., Buchholz, F., and Wang,
  Y.-M.}
\newblock Provenance-aware tracing of worm break-in and contaminations: A
  process coloring approach.
\newblock In {\em ICDCS\/} (2006).

\bibitem{mace}
{\sc Killian, C.~E., Anderson, J.~W., Braud, R., Jhala, R., and Vahdat, A.~M.}
\newblock Mace: Language support for building distributed systems.
\newblock In {\em PLDI\/} (2007).

\bibitem{intrusionrecovery}
{\sc Kim, T., Wang, X., Zeldovich, N., and Kaashoek, M.~F.}
\newblock Intrusion recovery using selective re-execution.
\newblock In {\em OSDI\/} (2010).

\bibitem{backtracking}
{\sc King, S.~T., and Chen, P.~M.}
\newblock Backtracking intrusions.
\newblock In {\em SOSP\/} (2003).

\bibitem{backtracking2}
{\sc King, S.~T., Mao, Z.~M., Lucchetti, D.~G., and Chen, P.~M.}
\newblock Enriching intrusion alerts through multi-host causality.
\newblock In {\em NDSS\/} (2005).

\bibitem{networkids}
{\sc Ko, C., Ruschitzka, M., and Levitt, K.~N.}
\newblock Execution monitoring of security-critical programs in distributed
  systems: a specification-based approach.
\newblock In {\em IEEE S\&P\/} (1997).

\bibitem{malwaresystemcall}
{\sc Kolbitsch, C., Comparetti, P.~M., Kruegel, C., Kirda, E., Zhou, X., and
  Wang, X.}
\newblock Effective and efficient malware detection at the end host.
\newblock In {\em {USENIX Security}\/} (2009).

\bibitem{idsbook}
{\sc Kruegel, C., Valeur, F., and Vigna, G.}
\newblock {\em Intrusion Detection and Correlation - Challenges and Solutions},
  vol.~14 of {\em Advances in Information Security}.
\newblock Springer, 2005.

\bibitem{accessminer}
{\sc Lanzi, A., Balzarotti, D., Kruegel, C., Christodorescu, M., and Kirda, E.}
\newblock Accessminer: Using system-centric models for malware protection.
\newblock In {\em CCS\/} (2010).

\bibitem{lee2013high}
{\sc Lee, K.~H., Zhang, X., and Xu, D.}
\newblock High accuracy attack provenance via binary-based execution partition.
\newblock In {\em NDSS\/} (2013).

\bibitem{loggc}
{\sc Lee, K.~H., Zhang, X., and Xu, D.}
\newblock Loggc: Garbage collecting audit log.
\newblock In {\em CCS\/} (2013).

\bibitem{networklang}
{\sc Loo, B.~T., Condie, T., Garofalakis, M., Gay, D.~E., Hellerstein, J.~M.,
  Maniatis, P., Ramakrishnan, R., Roscoe, T., and Stoica, I.}
\newblock Declarative networking: Language, execution and optimization.
\newblock In {\em SIGMOD\/} (2006).

\bibitem{Ma:2015:ALC:2818000.2818039}
{\sc Ma, S., Lee, K.~H., Kim, C.~H., Rhee, J., Zhang, X., and Xu, D.}
\newblock Accurate, low cost and instrumentation-free security audit logging
  for windows.
\newblock In {\em ACSAC\/} (2015).

\bibitem{mpi}
{\sc Ma, S., Zhai, J., Wang, F., Lee, K.~H., Zhang, X., and Xu, D.}
\newblock {MPI}: Multiple perspective attack investigation with semantic aware
  execution partitioning.
\newblock In {\em {USENIX} Security\/} (2017).

\bibitem{protracer}
{\sc Ma, S., Zhang, X., and Xu, D.}
\newblock Protracer: Towards practical provenance tracing by alternating
  between logging and tainting.
\newblock In {\em NDSS\/} (2016).

\bibitem{insider}
{\sc Senator, T.~E., Goldberg, H.~G., Memory, A., Young, W.~T., Rees, B.,
  Pierce, R., Huang, D., Reardon, M., Bader, D.~A., Chow, E., Essa, I., Jones,
  J., Bettadapura, V., Chau, D.~H., Green, O., Kaya, O., Zakrzewska, A.,
  Briscoe, E., Mappus, R. I.~L., McColl, R., Weiss, L., Dietterich, T.~G.,
  Fern, A., Wong, W.-K., Das, S., Emmott, A., Irvine, J., Lee, J.-Y., Koutra,
  D., Faloutsos, C., Corkill, D., Friedland, L., Gentzel, A., and Jensen, D.}
\newblock Detecting insider threats in a real corporate database of computer
  usage activity.
\newblock In {\em KDD\/} (2013).

\bibitem{backtrackingfile}
{\sc Sitaraman, S., and Venkatesan, S.}
\newblock Forensic analysis of file system intrusions using improved
  backtracking.
\newblock In {\em IWIA\/} (2005).

\bibitem{tquel}
{\sc Snodgrass, R.}
\newblock The temporal query language tquel.
\newblock {\em TODS 12}, 2 (1987), 247--298.

\bibitem{timesql}
{\sc Snodgrass, R.~T.}
\newblock {\em Developing Time-oriented Database Applications in SQL}.
\newblock Morgan Kaufmann Publishers Inc., 2000.

\bibitem{networkids2}
{\sc Sommer, R., and Paxson, V.}
\newblock Outside the closed world: On using machine learning for network
  intrusion detection.
\newblock In {\em IEEE S\&P\/} (2010).

\bibitem{Sommer:2014:HAE:2663716.2663735}
{\sc Sommer, R., Vallentin, M., De~Carli, L., and Paxson, V.}
\newblock Hilti: An abstract execution environment for deep, stateful network
  traffic analysis.
\newblock In {\em IMC\/} (2014).

\bibitem{fileanomaly}
{\sc Stolfo, S.~J., Hershkop, S., Bui, L.~H., Ferster, R., and Wang, K.}
\newblock Anomaly detection in computer security and an application to file
  system accesses.
\newblock In {\em ISMIS\/} (2005).

\bibitem{staticanalyzer}
{\sc Sung, A.~H., Xu, J., Chavez, P., and Mukkamala, S.}
\newblock Static analyzer of vicious executables ({SAVE}).
\newblock In {\em ACSAC\/} (2004).

\bibitem{Vallentin:2016:VUP:2930611.2930634}
{\sc Vallentin, M., Paxson, V., and Sommer, R.}
\newblock Vast: A unified platform for interactive network forensics.
\newblock In {\em NSDI\/} (2016).

\bibitem{reduction}
{\sc Xu, Z., Wu, Z., Li, Z., Jee, K., Rhee, J., Xiao, X., Xu, F., Wang, H., and
  Jiang, G.}
\newblock High fidelity data reduction for big data security dependency
  analyses.
\newblock In {\em CCS\/} (2016).

\bibitem{trafficaggregation}
{\sc Yen, T.-F., and Reiter, M.~K.}
\newblock Traffic aggregation for malware detection.
\newblock In {\em DIMVA\/} (2008).

\bibitem{securitylandslide}
{\sc Yu, S.}
\newblock Understanding the security vendor landscape using the cyber defense
  matrix, 2016.
\newblock RSA Conferences.

\bibitem{networkmalware}
{\sc Zhang, H., Yao, D.~D., and Ramakrishnan, N.}
\newblock Detection of stealthy malware activities with traffic causality and
  scalable triggering relation discovery.
\newblock In {\em ASIA CCS\/} (2014).

\end{thebibliography}

%\theendnotes

%\input{appendix}
\end{document}